\documentclass[10pt]{article}
\usepackage{amssymb}
\begin{document}
\begin{center}{\Large {\bf Ising Metamagnet Driven by Propagating Magnetic 
Field Wave:}}\\
{\Large {\bf Nonequilibrium Phases and Transitions}}\end{center}

\vskip 1 cm

\begin{center}{\it Muktish Acharyya}\\
{\it Department of Physics, Presidency University,}\\
{\it 86/1 College Street, Calcutta-700073, India.}\\
{E-mail:muktish.physics@presiuniv.ac.in}\end{center}

\vskip 1 cm

\noindent {\bf Abstract:} The nonequilibrium responses of Ising metamagnet
(layered antiferromagnet) to the propagating magnetic wave are studied by
Monte Carlo simulation. Here, the spatio-temporal variations of magnetic
field keeps the system away from equilibrium.
The sublattice magnetisations show dynamical symmetry
breaking in the low temperature ordered phase. The nonequilibrium phase 
transitions are studied from the temperature dependences of dynamic staggered
order parameter, its derivative and the dynamic specific heat. 
The transitions are marked by the peak position of dynamic specific heat and
the position of dip of the derivative of dynamic staggered order parameter.
It is observed that, for lower values of the amplitudes of the 
propagating magnetic field,
if the system is cooled from a high temperature, it undergoes a phase 
transition showing sharp peak of dynamic specific heat and sharp dip of the
derivative of the dynamic staggered order parameter. However, for higher
values of the amplitude of the propagating magnetic field, system 
exhibits multiple phase transitions. A comprehensive phase diagram is also
plotted. The transition, for vanishingly small value of the amplitude of the
propagating wave, is very close to that for 
equilibrium ferro-para phase transition of
cubic Ising ferromagnet.

\vskip 1 cm


\newpage

\noindent {\bf I. Introduction:}

The nonequilibrium responses of Ising ferromagnet to a time dependent magnetic field
give rise to several interesting phenomena \cite{rev}. Apart from dynamic hysteresis,
the dynamic phase transition is an emerging field of modern research. An oscillating
(in time) and uniform (in space) magnetic field yields dynamic phase transition accompanied
by dynamic symmetry breaking. 
Very recently, \cite{nucl,phs,br,pol,rfim} the responses of
Ising ferromagnet to a magnetic field having spatio-temporal variation, have become an interesting
subject of investigations. The classical nucleation phenomenon has been studied \cite{nucl}
in Ising ferromagnet in presence of a field having spatio-temporal (spreading Gaussian) variation.
Here, existence of a new time scale was mentioned. The nonequilibrium phase transition is
studied \cite{phs,pol} in Ising ferromagnet driven by polarised magnetic field. Multiple
nonequilibrium phase transitions were observed
\cite{br} in Ising ferromagnet irradiated by spherical magnetic wave. Even, the random
field Ising model exists in multiple nonequilibrium phases at zero temperature, 
if driven by plane polarised magnetic wave.

The real metamagnet, for example the iron chloride ${FeCl_2}$ show some interesting equilibrium
phases \cite{book}. The experiments \cite{expts1,expts2}
on iron bromide ${FeBr_2}$ show the existence of multiple
phase transitions (in the high field region). Several attempts have been made to understand this
peculiar phase diagram, theoretically \cite{theo1} in Ising metamagnets. However, this multiple
phase transition (indicated by double peak in specific heat) was successfully reproduced
\cite{dui} in Heisenberg metamagnet using Monte Carlo simulation. 
 
The phase transitions mentioned above the metamagnetic systems are of equilibrium type. Due
to the complicated interactions, metamagnets may exhibit interesting nonequilibrium phase
transition. This prompted the researchers to study the dynamic phase transition in metamagnets
in the presence of oscillating (in time only) magnetic field \cite{theo2}, in meanfield
approximation. The Monte Carlo study \cite{meta} was done in Ising metamagnet driven by sinusoidally
oscillating magnetic field. The nonequilibruim phase transition is studied and a phase boundary
is drawn in the plane formed by the temperature and amplitude of the oscillating magnetic field.
%%% New in Revised version%
Very recently, the nonequilibrium behaviours of synthetic metamagnetic 
(ising type) film are
studied \cite{pleimling} by Monte Carlo simulation. The time dependent response,
growth of domains inside the material and surface autocorrelation
(for thick films) are studied
in details. The phase diagrams are also drawn in field-temperature plane.
However, the behaviours of metamagnets are not yet investigated for a 
field having spatio-temporal variation.

In this paper, the responses of Ising metamagnet to a plane polarised magnetic wave
(the magnetic field has spatio-temporal variation), are studied by Monte Carlo simulation.
The paper is organised as follows: In the next section the model and the simulation
scheme are described. Section III contains the numerical results and the paper ends with
summary in section IV. 

\vskip 1cm

\noindent {\bf II. Model and Simulation:} 

The Ising metamagnet (layered antiferromagnet) can be modelled by the following Hamiltonian:

\begin{equation}
H = -{J_F} \sum s(x,y,z,t)s(x',y',z,t) - {J_{AF}} \sum s(x,y,z,t)s(x,y,z',t) - \sum h(y,t)s(x,y,z,t) 
\end{equation}

Where, $s(x,y,z,t)=\pm 1$, represents the Ising spins in the position $(x,y,z)$ at the time
instant $t$. First term represents the nearest neighbour
ferromagnetic ($J_F > 0$) interaction in a particular (xy) plane (same $z$). Second term represents
the nearest neighbour antiferromagnetic interaction ($J_{AF} < 0$) between two 
adjacent planes. 
Third term represents the spin-field interaction. $h(y,t)$ represents the value of magnetic
field at position $y$ at time $t$, due to the propagation of magnetic wave, 
propagating along the $y$ direction. The magnetic field of propagating 
(along the $y$ direction) magnetic wave
is represented as:

\begin{equation}
h(y,t) = h_0 {\rm cos} 2\pi (ft - y/{\lambda})
\end{equation}

where, $h_0$, $f$ and $\lambda$ denote the amplitude, frequency and wavelength of the
propagating magnetic wave respectively. 
The wavefront is parallel to $xz$ plane.
The model is defined in a cube of linear 
size $L$. The periodic boundary conditions are applied in all $(x,y,z)$ directions.

This model Ising metamagnet irradiated by plane magnetic wave is studied by Monte
Carlo simulation. The initial random spin configuration was taken such that statistically half
of the total number of spins has value +1. This corresponds to very high temperature
configuration. The nonequilibrium steady state at any given temperature was achieved by cooling
from high temperature configuration. At any given temperature the steady state configuration
was obtained by single spin 
(chosen randomly) flip Metropolis algorithm\cite{binder}. Where the probability of spin flip
is given by:

\begin{equation}
P_M (s(x,y,z,t) \to -s(x,y,z,t)) = {\rm Min} [1, {\rm exp}(-{{\Delta H} \over {kT}})]
\end{equation}

where, $\Delta H$ is change in energy due to spin flip. $k$ is the Boltzmann constant and 
$T$ is the temperature of the system. $L\times L \times L$ number of such random updates
constitutes a single Monte Carlo Step per Spin (MCSS) and defines the unit of time in the
problem. At any fixed temperature ($T$), $2\times 10^5$ MCSS time was devoted to get the
nonequilibrium steady configuration. It was checked that this is sufficient. Initial,
$10^5$ MCSS was discarded and the quantities in NESS were calculated from next $10^5$ MCSS.
The slow cooling was assumed by lowering the temperature in small steps ($\Delta T=0.05$).

Here, $L=20$ is considered. The system can be divided by two sublattices. The odd planes
form sublattice of type $'a'$ (say) and the even planes form the $'b'$ type sublattice. The
instantaneous magnetisation in $a$ sublattice is denoted by 
$m_a = 2 \sum s(x,y,z,t)/{(L\times L \times L)}$ summing over odd planes only. Similarly, $m_b(t)$
can be defined by summing over the even planes only. The instantaneous 
value of the staggered magnetisation
is defined as $m_s(t) = (m_a(t)-m_b(t))/2$. 
The sublattice dynamic order parameters are defined as $Q_a = f \oint m_a(t) dt$ and
$Q_b = f \oint m_b(t) dt$.
The dynamic staggered order parameter is defined
as $Q_s = f \oint m_s(t) dt$. The dynamic energy of the system is defined as $E = f \oint H(t) dt$.
These are defined as the time averaged over the full cycle of the propagating magnetic field.
The dynamic specific heat is defined as $C = {{dE} \over {dT}}$.

%%%% Added in revised version
Here, the choice of the system size ($L=20$) is just a compromise between the available
computational facilities (Intel CORE i5 processor)
 and the computational time. 
A few results are checked in smaller sizes and no considerable qualitative changes
were found. In this simulation, since the spatio-temporal variation of the applied magnetic
field is considered, the updating of spins requires more time than that required for
steady field or oscillating (but uniform over the space) field. The CPU time required for
$2\times10^5$ MCSS is 13.5 Minutes.

The units of the parameters used in the simulation are specified as follows:.
the field amplitude $h_0$ is measured in the unit of $J_F$ and the temperature
$T$ is measured in the unit of $J_F/k$, the frequency $f$ is measured in the
unit of MCSS$^{-1}$ and the wavelength $\lambda$ is measured in the unit
of lattice spacing.

\vskip 1cm

\noindent {\bf III. Results:}

In the present study, the frequency ($f=0.01$) and the wavelength
($\lambda=5$) of the propagating magnetic field are kept constant.
The strengths of the interactions, $J_F = 1.0$ and $J_{AF} = -J_F$ were also
kept constant throughout the study.
The instantaneous sublattice magnetisations $m_a(t)$ and $m_b(t)$ are studied for two different
temperatures $T = 4.5$ and $T=3.0$ for fixed $h_0=2.0$. These plots are shown in Fig-1. In the
high temperature ($T=4.5$), both $m_a(t)$ (in Fig-1(a)) and $m_b(t)$ (in Fig-1(b)) oscillates
symmetrically (around zero magnetisation). As a result, the staggered magnetisation would also
oscillate symmetrically. These correspond to the dynamically symmetric phase. On the other hand,
for lower temperature ($T=3.0$ here), both oscillate asymmetrically (around zero magnetisation).
These are shown in Fig-1(c) and Fig-1(d). Consequently, the staggered magnetisation $m_s(t)$,
also oscillates asymmetrically. This is symmetry broken dynamically ordered phase. The dynamic
staggered order parameter $Q_s$ would be zero in the symmetric disordered 
phase and becomes nonzero in
the low temperature dynamically ordered symmetry broken phase.

The temperature dependences of all dynamic quantities 
($Q_a$, $Q_b$, $Q_s$, ${{dQ_s} \over {dT}}$, $E$ and $C$) are shown in Fig-2. Here, all these
dynamic quantities are studied 
as a function of temperature ($T$) for two values of field amplitudes ($h_0 = 1.0$ and
$h_0 = 3.0$). As the temperature is lowered from a high temperature the staggered dynamic order
parameters ($Q_a$ and $Q_b$) are observed to become nonzero (getting the ordered phase) near
the dynamic transition temperature. These transitions are observed (Fig-2(a)) to happen at higher
temperatures for lower value of field amplitude ($h_0$). The dynamic staggered order parameter
also becomes nonzero near the dynamic transition point. The transition temperatures (Fig-2(b))
are observed to be lowered for higher value of field amplitude. The dynamic staggered order
parameter $Q_s$ is assumed to behave like $Q_s \sim (T_c - T)^{\beta}$, (where $\beta < 1$).
Keeping this in mind, the derivative, ${{dQ_s} \over {dT}}$ is studied as function of temperature
for two different values of field amplitudes (mentioned above). These show very sharp dips
(eventually divergences for $L \to \infty$) near the transition points (Fig-2(c)). Actually,
the transitions are well marked and the transition temperatures are estimated from the positions
of the dips of ${{dQ_s} \over {dT}}$. Here also, the dips form at lower temperatures for higher
values of the field amplitudes and vice versa. The temperature dependences of dynamic energy
$E$, are shown in Fig-2(d). The inflection point acts as the marker of dynamic phase transition.
In this case, it is clearly observed that these inflection points shifts towards the lower
temperature for higher value of the magnetic field. The most physical quantity to detect the
phase transition is specific heat. Here, in Fig-2(e), the temperature dependence of dynamic
specific heat $C$ is shown. The sharp peak of specific heat indicates the dynamic transition
point. The transitions were observed to happen at lower temperatures for higher field 
amplitudes. From these studies, it is observed that the dynamic transition temperature is a
function of field amplitude of the propagating magnetic field. For the lower values of the
field amplitudes (mentioned above) the dynamic transitions, observed here are single transition
(indicated by single peak of $C$ or single dip of ${{dQ_s} \over {dT}}$).

However, for higher values of field amplitudes, e.g., $h_0 = 3.9$ and $h_0 = 4.0$. The multiple
(actually double) dynamic transitions are observed. This is quite interesting and already 
observed in real and model metamagnets for constant (in time) and uniform (in space) magnetic
field. Similar (mentioned in last para) study is done and shown in Fig-3. Here, in addition to
high temperature sharp dip of ${{dQ_s} \over {dT}}$, a smeared dip was found at lower temperature.
This indicates another transition (Fig-3(c)). The same was observed for $C$. Two peaks of 
dynamic specific heat indicates two dynamic transitions (Fig-3(e)). As the value of field
amplitudes increases the transitions occurs at lower temperatures.

The dynamic transition temperatures are estimated from the dips of ${{dQ_s} \over {dT}}$ and
peaks of $C$ for different values of $h_0$. It is observed that the transition temperature is
function of $h_0$ or vice versa. Collecting all data of dynamic transition temperatures for
different values of field amplitudes, the comprehensive phase diagram is plotted and shown
in Fig-4. By cooling the system from high temperature (for lower values of 
amplitude of the propagating magnetic field) one gets the single
phase transition (boundary marked by the solid symbols in Fig-4).
The transition occurs at higher temperatures as the value of amplitude of the
propagating field decreases. In the limit of vanishingly small amplitude the
transition occurs at a temperature which is very close to the normal ferro-para
transition temperature ($T_c = 4.511...$) of cubic Ising ferromagnet.
On the other
hand, for higher values of amplitude of propagating field, the system exhibits
two phase transitions. The high temperature transition (marked by solid
symbols) and low temperature transition (marked by open symbols).  

\vskip 1cm

\noindent {\bf IV. Summary:}

In this paper, the dynamical responses of Ising metamagnet (layered 
antiferromagnet) to the plane propagating magnetic wave, are studied by Monte
Carlo simulation. The interaction between the magnetic field (space and time
dependent) of the propagating
wave and the Ising spins, keeps the system away from the equilibrium.

The time variations of the sublattice magnetisations are
studied. For a fixed set of values of field amplitude, frequency and 
wavelength, a dynamical symmetry breaking is observed as one cools the system
from a high temperature. This dynamic symmetry breaking is also related to a
nonequilibrium phase transition. The nonequilibrium phase transition is 
studied by the temperature variations of dynamic staggered order parameter,
its derivative and the dynamic specific heat. The transition temperatures
are estimated from the positions of peaks of dynamic specific heat and the
positions of dips of the derivative of the dynamic staggered order
parameters.

For lower values of the amplitudes of the propagating magnetic field,
if the system is cooled from a high temperature, it undergoes a phase 
transition showing sharp peak of dynamic specific heat and sharp dip of the
derivative of the dynamic staggered order parameter. However, for higher
values of the amplitude of the propagating magnetic field, system 
exhibits multiple phase transitions. A comprehensive phase diagram is also
plotted in the plane formed by the temperature and the amplitude of the 
propagating magnetic field wave.

The comprehensive phase boundary obtained here may be realised 
qualitatively and physically
as follows \cite{pleimling}: for lower values of 
field amplitude the system undergoes a 
transition from paramagnetic to antiferromagnetic (layered) phase as the
temperetured is lowered, since the 
intra layer antiferromagnetic interaction strength wins over the
competition against the field strength. 
As
the strength of the field increases, it becomes relatively stronger enough
to create
the clusters of up spins within the plane of down spins. As a result, a
mixed phase may appear. As a result, the double transition is oberseved
for higher values of field amplitude.

In this problem, several important observations are yet to be made. 
For example, the size of the system used in this study, is not yet confirmed
to be sufficient for the conclusions based upon the observations. The orders
of the phase transitions and a possible existence of any tricritical point
on the phase boundary may be an interesting study. For that purpose, the
studies of distributions of order parameters and the temperature variation
of Binder cumulant, would be the useful trategy.
The
finite size analysis is one important issue. The dependences of the phase
boundary on the wavelengths, frequency and the antiferromagnetic interaction
strength would also be very interesting study. The work in this front is 
going on. It would be interesting if the reported results agree well with the
expertiments.

\vskip 1cm

\noindent {\bf Acknowledgements:} The library facilities provided by 
the Calcutta University is gratefully acknowledged.

\newpage

% GNUPLOT: LaTeX picture FIG-1
\setlength{\unitlength}{0.240900pt}
\ifx\plotpoint\undefined\newsavebox{\plotpoint}\fi
\sbox{\plotpoint}{\rule[-0.200pt]{0.400pt}{0.400pt}}%


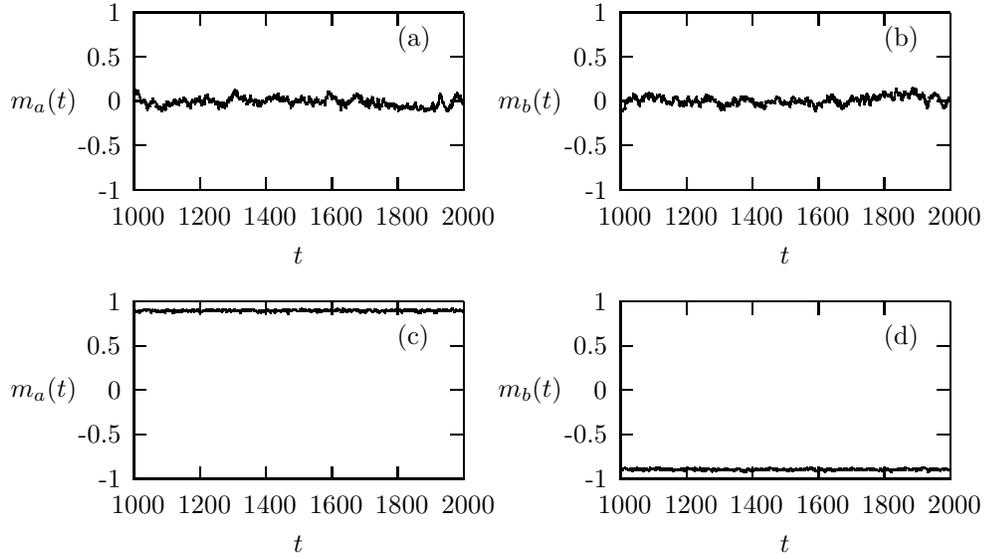
\noindent {\bf Fig-1.} The time evolution of sublattice magnetisations in
the nonequilibrium steady state. (a) and (b) show the dynamically symmetric
disordered phase at $h_0=2.0$ and $T=4.50$. (c) and (d) show the dynamically
symmetry broken ordered phase at $h_0=2.0$ and $T=3.0$. 
Here, $f=0.01$ and $\lambda = 5$.
\newpage

% GNUPLOT: LaTeX picture FIG-2
\setlength{\unitlength}{0.240900pt}
\ifx\plotpoint\undefined\newsavebox{\plotpoint}\fi
\sbox{\plotpoint}{\rule[-0.200pt]{0.400pt}{0.400pt}}%
%

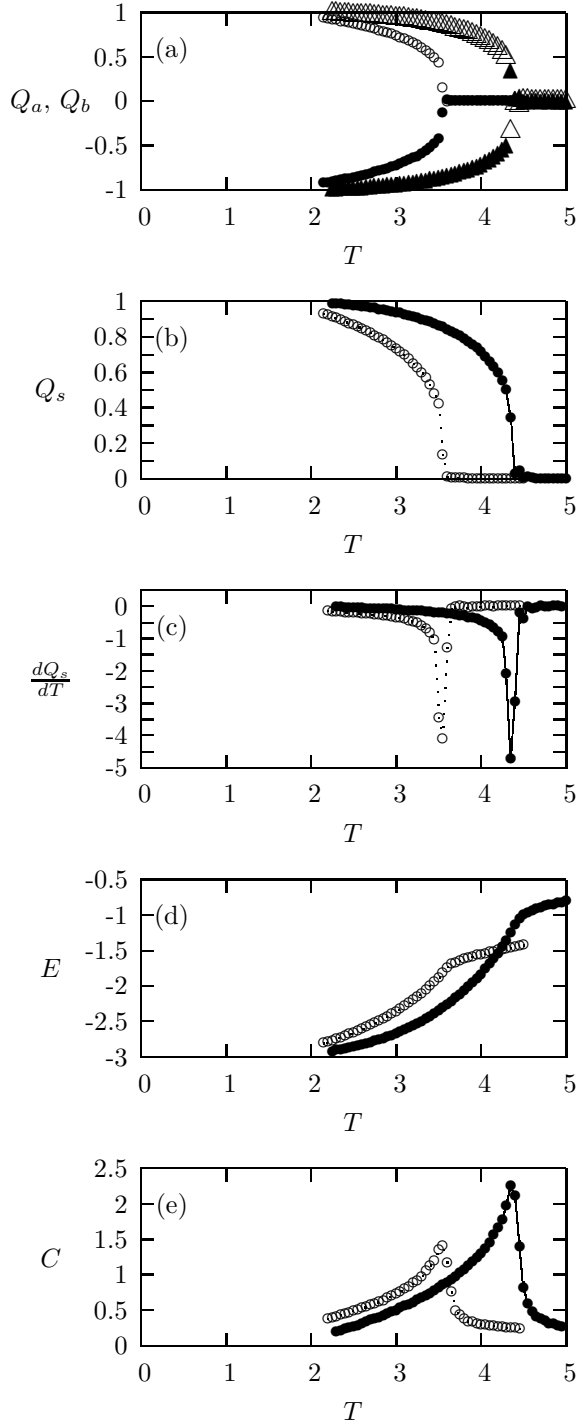
\noindent {\bf Fig-2.} Temperature dependences of all dynamic quantities.
(a) $Q_a (\circ)$ and $Q_b (\bullet)$ for $h_0=3.0$ and $Q_a (\triangle)$ and
$Q_b (\blacktriangle)$ for $h_0=1.0$. (b) $Q_s$ versus $T$ for $h_0 = 3.0
(\circ)$ and $h_0 = 1.0 (\bullet)$. (c) ${{dQ_s} \over {dT}}$ versus $T$, for
$h_0 = 3.0 (\circ)$ and $h_0 = 1.0 (\bullet)$. (d) $E$ versus $T$, for
$h_0 = 3.0 (\circ)$ and $h_0 = 1.0 (\bullet)$. (e) $C = {{dE} \over {dT}}$ 
versus $T$, for $h_0 = 3.0 (\circ)$ and $h_0 = 1.0 (\bullet)$.
Here, $f=0.01$ and $\lambda = 5$.

\newpage

% GNUPLOT: LaTeX picture FIG-3
\setlength{\unitlength}{0.240900pt}
\ifx\plotpoint\undefined\newsavebox{\plotpoint}\fi
\sbox{\plotpoint}{\rule[-0.200pt]{0.400pt}{0.400pt}}%
\begin{picture}(900,450)(0,0)
\sbox{\plotpoint}{\rule[-0.200pt]{0.400pt}{0.400pt}}%
\put(171.0,131.0){\rule[-0.200pt]{4.818pt}{0.400pt}}
\put(151,131){\makebox(0,0)[r]{-1}}
\put(819.0,131.0){\rule[-0.200pt]{4.818pt}{0.400pt}}
\put(171.0,201.0){\rule[-0.200pt]{4.818pt}{0.400pt}}
\put(151,201){\makebox(0,0)[r]{-0.5}}
\put(819.0,201.0){\rule[-0.200pt]{4.818pt}{0.400pt}}
\put(171.0,271.0){\rule[-0.200pt]{4.818pt}{0.400pt}}
\put(151,271){\makebox(0,0)[r]{ 0}}
\put(819.0,271.0){\rule[-0.200pt]{4.818pt}{0.400pt}}
\put(171.0,340.0){\rule[-0.200pt]{4.818pt}{0.400pt}}
\put(151,340){\makebox(0,0)[r]{ 0.5}}
\put(819.0,340.0){\rule[-0.200pt]{4.818pt}{0.400pt}}
\put(171.0,410.0){\rule[-0.200pt]{4.818pt}{0.400pt}}
\put(151,410){\makebox(0,0)[r]{ 1}}
\put(819.0,410.0){\rule[-0.200pt]{4.818pt}{0.400pt}}
\put(171.0,131.0){\rule[-0.200pt]{0.400pt}{4.818pt}}
\put(171,90){\makebox(0,0){ 0}}
\put(171.0,390.0){\rule[-0.200pt]{0.400pt}{4.818pt}}
\put(305.0,131.0){\rule[-0.200pt]{0.400pt}{4.818pt}}
\put(305,90){\makebox(0,0){ 1}}
\put(305.0,390.0){\rule[-0.200pt]{0.400pt}{4.818pt}}
\put(438.0,131.0){\rule[-0.200pt]{0.400pt}{4.818pt}}
\put(438,90){\makebox(0,0){ 2}}
\put(438.0,390.0){\rule[-0.200pt]{0.400pt}{4.818pt}}
\put(572.0,131.0){\rule[-0.200pt]{0.400pt}{4.818pt}}
\put(572,90){\makebox(0,0){ 3}}
\put(572.0,390.0){\rule[-0.200pt]{0.400pt}{4.818pt}}
\put(705.0,131.0){\rule[-0.200pt]{0.400pt}{4.818pt}}
\put(705,90){\makebox(0,0){ 4}}
\put(705.0,390.0){\rule[-0.200pt]{0.400pt}{4.818pt}}
\put(839.0,131.0){\rule[-0.200pt]{0.400pt}{4.818pt}}
\put(839,90){\makebox(0,0){ 5}}
\put(839.0,390.0){\rule[-0.200pt]{0.400pt}{4.818pt}}
\put(171.0,131.0){\rule[-0.200pt]{0.400pt}{67.211pt}}
\put(171.0,131.0){\rule[-0.200pt]{160.921pt}{0.400pt}}
\put(839.0,131.0){\rule[-0.200pt]{0.400pt}{67.211pt}}
\put(171.0,410.0){\rule[-0.200pt]{160.921pt}{0.400pt}}
\put(30,270){\makebox(0,0){$Q_a$, $Q_b$}}
\put(220,350){\makebox(0,0){(a)}}
\put(505,29){\makebox(0,0){$T$}}
\put(572,270){\makebox(0,0){$\bullet$}}
\put(565,270){\makebox(0,0){$\bullet$}}
\put(558,270){\makebox(0,0){$\bullet$}}
\put(552,270){\makebox(0,0){$\bullet$}}
\put(545,271){\makebox(0,0){$\bullet$}}
\put(538,271){\makebox(0,0){$\bullet$}}
\put(532,271){\makebox(0,0){$\bullet$}}
\put(525,271){\makebox(0,0){$\bullet$}}
\put(518,271){\makebox(0,0){$\bullet$}}
\put(512,270){\makebox(0,0){$\bullet$}}
\put(505,270){\makebox(0,0){$\bullet$}}
\put(498,269){\makebox(0,0){$\bullet$}}
\put(492,274){\makebox(0,0){$\bullet$}}
\put(485,277){\makebox(0,0){$\bullet$}}
\put(478,262){\makebox(0,0){$\bullet$}}
\put(472,292){\makebox(0,0){$\bullet$}}
\put(465,222){\makebox(0,0){$\bullet$}}
\put(458,218){\makebox(0,0){$\bullet$}}
\put(452,215){\makebox(0,0){$\bullet$}}
\put(445,213){\makebox(0,0){$\bullet$}}
\put(438,211){\makebox(0,0){$\bullet$}}
\put(432,209){\makebox(0,0){$\bullet$}}
\put(425,207){\makebox(0,0){$\bullet$}}
\put(418,205){\makebox(0,0){$\bullet$}}
\put(411,203){\makebox(0,0){$\bullet$}}
\put(405,202){\makebox(0,0){$\bullet$}}
\put(398,200){\makebox(0,0){$\bullet$}}
\put(391,199){\makebox(0,0){$\bullet$}}
\put(385,197){\makebox(0,0){$\bullet$}}
\put(378,196){\makebox(0,0){$\bullet$}}
\put(371,194){\makebox(0,0){$\bullet$}}
\put(365,192){\makebox(0,0){$\bullet$}}
\put(358,190){\makebox(0,0){$\bullet$}}
\put(351,187){\makebox(0,0){$\bullet$}}
\put(345,185){\makebox(0,0){$\bullet$}}
\put(338,182){\makebox(0,0){$\bullet$}}
\put(331,179){\makebox(0,0){$\bullet$}}
\put(325,175){\makebox(0,0){$\bullet$}}
\put(318,171){\makebox(0,0){$\bullet$}}
\put(311,165){\makebox(0,0){$\bullet$}}
\put(305,160){\makebox(0,0){$\bullet$}}
\put(298,154){\makebox(0,0){$\bullet$}}
\put(291,148){\makebox(0,0){$\bullet$}}
\put(285,143){\makebox(0,0){$\bullet$}}
\put(278,138){\makebox(0,0){$\bullet$}}
\put(271,135){\makebox(0,0){$\bullet$}}
\put(265,133){\makebox(0,0){$\bullet$}}
\put(258,132){\makebox(0,0){$\bullet$}}
\put(572,271){\makebox(0,0){$\circ$}}
\put(565,271){\makebox(0,0){$\circ$}}
\put(558,271){\makebox(0,0){$\circ$}}
\put(552,271){\makebox(0,0){$\circ$}}
\put(545,270){\makebox(0,0){$\circ$}}
\put(538,270){\makebox(0,0){$\circ$}}
\put(532,270){\makebox(0,0){$\circ$}}
\put(525,270){\makebox(0,0){$\circ$}}
\put(518,270){\makebox(0,0){$\circ$}}
\put(512,271){\makebox(0,0){$\circ$}}
\put(505,271){\makebox(0,0){$\circ$}}
\put(498,272){\makebox(0,0){$\circ$}}
\put(492,267){\makebox(0,0){$\circ$}}
\put(485,264){\makebox(0,0){$\circ$}}
\put(478,279){\makebox(0,0){$\circ$}}
\put(472,249){\makebox(0,0){$\circ$}}
\put(465,319){\makebox(0,0){$\circ$}}
\put(458,323){\makebox(0,0){$\circ$}}
\put(452,326){\makebox(0,0){$\circ$}}
\put(445,328){\makebox(0,0){$\circ$}}
\put(438,330){\makebox(0,0){$\circ$}}
\put(432,332){\makebox(0,0){$\circ$}}
\put(425,334){\makebox(0,0){$\circ$}}
\put(418,336){\makebox(0,0){$\circ$}}
\put(411,338){\makebox(0,0){$\circ$}}
\put(405,339){\makebox(0,0){$\circ$}}
\put(398,341){\makebox(0,0){$\circ$}}
\put(391,342){\makebox(0,0){$\circ$}}
\put(385,344){\makebox(0,0){$\circ$}}
\put(378,346){\makebox(0,0){$\circ$}}
\put(371,347){\makebox(0,0){$\circ$}}
\put(365,349){\makebox(0,0){$\circ$}}
\put(358,351){\makebox(0,0){$\circ$}}
\put(351,354){\makebox(0,0){$\circ$}}
\put(345,356){\makebox(0,0){$\circ$}}
\put(338,359){\makebox(0,0){$\circ$}}
\put(331,362){\makebox(0,0){$\circ$}}
\put(325,366){\makebox(0,0){$\circ$}}
\put(318,370){\makebox(0,0){$\circ$}}
\put(311,375){\makebox(0,0){$\circ$}}
\put(305,381){\makebox(0,0){$\circ$}}
\put(298,387){\makebox(0,0){$\circ$}}
\put(291,393){\makebox(0,0){$\circ$}}
\put(285,398){\makebox(0,0){$\circ$}}
\put(278,403){\makebox(0,0){$\circ$}}
\put(271,406){\makebox(0,0){$\circ$}}
\put(265,408){\makebox(0,0){$\circ$}}
\put(258,409){\makebox(0,0){$\circ$}}
\sbox{\plotpoint}{\rule[-0.400pt]{0.800pt}{0.800pt}}%
\put(639,271){\makebox(0,0){$\triangle$}}
\put(632,270){\makebox(0,0){$\triangle$}}
\put(625,271){\makebox(0,0){$\triangle$}}
\put(619,271){\makebox(0,0){$\triangle$}}
\put(612,270){\makebox(0,0){$\triangle$}}
\put(605,271){\makebox(0,0){$\triangle$}}
\put(599,271){\makebox(0,0){$\triangle$}}
\put(592,271){\makebox(0,0){$\triangle$}}
\put(585,271){\makebox(0,0){$\triangle$}}
\put(578,270){\makebox(0,0){$\triangle$}}
\put(572,270){\makebox(0,0){$\triangle$}}
\put(565,271){\makebox(0,0){$\triangle$}}
\put(558,270){\makebox(0,0){$\triangle$}}
\put(552,271){\makebox(0,0){$\triangle$}}
\put(545,270){\makebox(0,0){$\triangle$}}
\put(538,271){\makebox(0,0){$\triangle$}}
\put(532,270){\makebox(0,0){$\triangle$}}
\put(525,268){\makebox(0,0){$\triangle$}}
\put(518,277){\makebox(0,0){$\triangle$}}
\put(512,277){\makebox(0,0){$\triangle$}}
\put(505,266){\makebox(0,0){$\triangle$}}
\put(498,322){\makebox(0,0){$\triangle$}}
\put(492,326){\makebox(0,0){$\triangle$}}
\put(485,330){\makebox(0,0){$\triangle$}}
\put(478,333){\makebox(0,0){$\triangle$}}
\put(472,336){\makebox(0,0){$\triangle$}}
\put(465,339){\makebox(0,0){$\triangle$}}
\put(458,341){\makebox(0,0){$\triangle$}}
\put(452,344){\makebox(0,0){$\triangle$}}
\put(445,346){\makebox(0,0){$\triangle$}}
\put(438,348){\makebox(0,0){$\triangle$}}
\put(432,350){\makebox(0,0){$\triangle$}}
\put(425,352){\makebox(0,0){$\triangle$}}
\put(418,354){\makebox(0,0){$\triangle$}}
\put(411,356){\makebox(0,0){$\triangle$}}
\put(405,358){\makebox(0,0){$\triangle$}}
\put(398,360){\makebox(0,0){$\triangle$}}
\put(391,362){\makebox(0,0){$\triangle$}}
\put(385,364){\makebox(0,0){$\triangle$}}
\put(378,367){\makebox(0,0){$\triangle$}}
\put(371,369){\makebox(0,0){$\triangle$}}
\put(365,372){\makebox(0,0){$\triangle$}}
\put(358,375){\makebox(0,0){$\triangle$}}
\put(351,378){\makebox(0,0){$\triangle$}}
\put(345,382){\makebox(0,0){$\triangle$}}
\put(338,386){\makebox(0,0){$\triangle$}}
\put(331,389){\makebox(0,0){$\triangle$}}
\put(325,393){\makebox(0,0){$\triangle$}}
\put(318,396){\makebox(0,0){$\triangle$}}
\put(311,400){\makebox(0,0){$\triangle$}}
\put(305,402){\makebox(0,0){$\triangle$}}
\put(298,405){\makebox(0,0){$\triangle$}}
\put(291,406){\makebox(0,0){$\triangle$}}
\put(285,408){\makebox(0,0){$\triangle$}}
\put(278,409){\makebox(0,0){$\triangle$}}
\put(271,409){\makebox(0,0){$\triangle$}}
\put(265,410){\makebox(0,0){$\triangle$}}
\put(258,410){\makebox(0,0){$\triangle$}}
\sbox{\plotpoint}{\rule[-0.500pt]{1.000pt}{1.000pt}}%
\put(639,270){\makebox(0,0){$\blacktriangle$}}
\put(632,271){\makebox(0,0){$\blacktriangle$}}
\put(625,270){\makebox(0,0){$\blacktriangle$}}
\put(619,270){\makebox(0,0){$\blacktriangle$}}
\put(612,271){\makebox(0,0){$\blacktriangle$}}
\put(605,270){\makebox(0,0){$\blacktriangle$}}
\put(599,270){\makebox(0,0){$\blacktriangle$}}
\put(592,270){\makebox(0,0){$\blacktriangle$}}
\put(585,270){\makebox(0,0){$\blacktriangle$}}
\put(578,271){\makebox(0,0){$\blacktriangle$}}
\put(572,271){\makebox(0,0){$\blacktriangle$}}
\put(565,270){\makebox(0,0){$\blacktriangle$}}
\put(558,271){\makebox(0,0){$\blacktriangle$}}
\put(552,270){\makebox(0,0){$\blacktriangle$}}
\put(545,271){\makebox(0,0){$\blacktriangle$}}
\put(538,270){\makebox(0,0){$\blacktriangle$}}
\put(532,271){\makebox(0,0){$\blacktriangle$}}
\put(525,273){\makebox(0,0){$\blacktriangle$}}
\put(518,264){\makebox(0,0){$\blacktriangle$}}
\put(512,264){\makebox(0,0){$\blacktriangle$}}
\put(505,275){\makebox(0,0){$\blacktriangle$}}
\put(498,219){\makebox(0,0){$\blacktriangle$}}
\put(492,215){\makebox(0,0){$\blacktriangle$}}
\put(485,211){\makebox(0,0){$\blacktriangle$}}
\put(478,208){\makebox(0,0){$\blacktriangle$}}
\put(472,205){\makebox(0,0){$\blacktriangle$}}
\put(465,202){\makebox(0,0){$\blacktriangle$}}
\put(458,200){\makebox(0,0){$\blacktriangle$}}
\put(452,197){\makebox(0,0){$\blacktriangle$}}
\put(445,195){\makebox(0,0){$\blacktriangle$}}
\put(438,193){\makebox(0,0){$\blacktriangle$}}
\put(432,191){\makebox(0,0){$\blacktriangle$}}
\put(425,189){\makebox(0,0){$\blacktriangle$}}
\put(418,187){\makebox(0,0){$\blacktriangle$}}
\put(411,185){\makebox(0,0){$\blacktriangle$}}
\put(405,183){\makebox(0,0){$\blacktriangle$}}
\put(398,181){\makebox(0,0){$\blacktriangle$}}
\put(391,179){\makebox(0,0){$\blacktriangle$}}
\put(385,177){\makebox(0,0){$\blacktriangle$}}
\put(378,174){\makebox(0,0){$\blacktriangle$}}
\put(371,172){\makebox(0,0){$\blacktriangle$}}
\put(365,169){\makebox(0,0){$\blacktriangle$}}
\put(358,166){\makebox(0,0){$\blacktriangle$}}
\put(351,163){\makebox(0,0){$\blacktriangle$}}
\put(345,159){\makebox(0,0){$\blacktriangle$}}
\put(338,156){\makebox(0,0){$\blacktriangle$}}
\put(331,152){\makebox(0,0){$\blacktriangle$}}
\put(325,148){\makebox(0,0){$\blacktriangle$}}
\put(318,145){\makebox(0,0){$\blacktriangle$}}
\put(311,141){\makebox(0,0){$\blacktriangle$}}
\put(305,138){\makebox(0,0){$\blacktriangle$}}
\put(298,136){\makebox(0,0){$\blacktriangle$}}
\put(291,135){\makebox(0,0){$\blacktriangle$}}
\put(285,133){\makebox(0,0){$\blacktriangle$}}
\put(278,132){\makebox(0,0){$\blacktriangle$}}
\put(271,132){\makebox(0,0){$\blacktriangle$}}
\put(265,131){\makebox(0,0){$\blacktriangle$}}
\put(258,131){\makebox(0,0){$\blacktriangle$}}
\sbox{\plotpoint}{\rule[-0.200pt]{0.400pt}{0.400pt}}%
\put(171.0,131.0){\rule[-0.200pt]{0.400pt}{67.211pt}}
\put(171.0,131.0){\rule[-0.200pt]{160.921pt}{0.400pt}}
\put(839.0,131.0){\rule[-0.200pt]{0.400pt}{67.211pt}}
\put(171.0,410.0){\rule[-0.200pt]{160.921pt}{0.400pt}}
\end{picture}

\begin{picture}(900,450)(0,0)
\put(171.0,131.0){\rule[-0.200pt]{4.818pt}{0.400pt}}
\put(151,131){\makebox(0,0)[r]{ 0}}
\put(819.0,131.0){\rule[-0.200pt]{4.818pt}{0.400pt}}
\put(171.0,159.0){\rule[-0.200pt]{4.818pt}{0.400pt}}
%\put(151,159){\makebox(0,0)[r]{ 0.1}}
\put(819.0,159.0){\rule[-0.200pt]{4.818pt}{0.400pt}}
\put(171.0,187.0){\rule[-0.200pt]{4.818pt}{0.400pt}}
\put(151,187){\makebox(0,0)[r]{ 0.2}}
\put(819.0,187.0){\rule[-0.200pt]{4.818pt}{0.400pt}}
\put(171.0,215.0){\rule[-0.200pt]{4.818pt}{0.400pt}}
%\put(151,215){\makebox(0,0)[r]{ 0.3}}
\put(819.0,215.0){\rule[-0.200pt]{4.818pt}{0.400pt}}
\put(171.0,243.0){\rule[-0.200pt]{4.818pt}{0.400pt}}
\put(151,243){\makebox(0,0)[r]{ 0.4}}
\put(819.0,243.0){\rule[-0.200pt]{4.818pt}{0.400pt}}
\put(171.0,271.0){\rule[-0.200pt]{4.818pt}{0.400pt}}
%\put(151,271){\makebox(0,0)[r]{ 0.5}}
\put(819.0,271.0){\rule[-0.200pt]{4.818pt}{0.400pt}}
\put(171.0,298.0){\rule[-0.200pt]{4.818pt}{0.400pt}}
\put(151,298){\makebox(0,0)[r]{ 0.6}}
\put(819.0,298.0){\rule[-0.200pt]{4.818pt}{0.400pt}}
\put(171.0,326.0){\rule[-0.200pt]{4.818pt}{0.400pt}}
%\put(151,326){\makebox(0,0)[r]{ 0.7}}
\put(819.0,326.0){\rule[-0.200pt]{4.818pt}{0.400pt}}
\put(171.0,354.0){\rule[-0.200pt]{4.818pt}{0.400pt}}
\put(151,354){\makebox(0,0)[r]{ 0.8}}
\put(819.0,354.0){\rule[-0.200pt]{4.818pt}{0.400pt}}
\put(171.0,382.0){\rule[-0.200pt]{4.818pt}{0.400pt}}
%\put(151,382){\makebox(0,0)[r]{ 0.9}}
\put(819.0,382.0){\rule[-0.200pt]{4.818pt}{0.400pt}}
\put(171.0,410.0){\rule[-0.200pt]{4.818pt}{0.400pt}}
\put(151,410){\makebox(0,0)[r]{ 1}}
\put(819.0,410.0){\rule[-0.200pt]{4.818pt}{0.400pt}}
\put(171.0,131.0){\rule[-0.200pt]{0.400pt}{4.818pt}}
\put(171,90){\makebox(0,0){ 0}}
\put(171.0,390.0){\rule[-0.200pt]{0.400pt}{4.818pt}}
\put(305.0,131.0){\rule[-0.200pt]{0.400pt}{4.818pt}}
\put(305,90){\makebox(0,0){ 1}}
\put(305.0,390.0){\rule[-0.200pt]{0.400pt}{4.818pt}}
\put(438.0,131.0){\rule[-0.200pt]{0.400pt}{4.818pt}}
\put(438,90){\makebox(0,0){ 2}}
\put(438.0,390.0){\rule[-0.200pt]{0.400pt}{4.818pt}}
\put(572.0,131.0){\rule[-0.200pt]{0.400pt}{4.818pt}}
\put(572,90){\makebox(0,0){ 3}}
\put(572.0,390.0){\rule[-0.200pt]{0.400pt}{4.818pt}}
\put(705.0,131.0){\rule[-0.200pt]{0.400pt}{4.818pt}}
\put(705,90){\makebox(0,0){ 4}}
\put(705.0,390.0){\rule[-0.200pt]{0.400pt}{4.818pt}}
\put(839.0,131.0){\rule[-0.200pt]{0.400pt}{4.818pt}}
\put(839,90){\makebox(0,0){ 5}}
\put(839.0,390.0){\rule[-0.200pt]{0.400pt}{4.818pt}}
\put(171.0,131.0){\rule[-0.200pt]{0.400pt}{67.211pt}}
\put(171.0,131.0){\rule[-0.200pt]{160.921pt}{0.400pt}}
\put(839.0,131.0){\rule[-0.200pt]{0.400pt}{67.211pt}}
\put(171.0,410.0){\rule[-0.200pt]{160.921pt}{0.400pt}}
\put(30,270){\makebox(0,0){$Q_s$}}
\put(220,350){\makebox(0,0){(b)}}
\put(505,29){\makebox(0,0){$T$}}
\put(639,131){\usebox{\plotpoint}}
\put(552,130.67){\rule{1.445pt}{0.400pt}}
\multiput(555.00,130.17)(-3.000,1.000){2}{\rule{0.723pt}{0.400pt}}
\put(558.0,131.0){\rule[-0.200pt]{19.513pt}{0.400pt}}
\put(538,131.67){\rule{1.686pt}{0.400pt}}
\multiput(541.50,131.17)(-3.500,1.000){2}{\rule{0.843pt}{0.400pt}}
\put(532,131.67){\rule{1.445pt}{0.400pt}}
\multiput(535.00,132.17)(-3.000,-1.000){2}{\rule{0.723pt}{0.400pt}}
\multiput(529.26,132.59)(-0.710,0.477){7}{\rule{0.660pt}{0.115pt}}
\multiput(530.63,131.17)(-5.630,5.000){2}{\rule{0.330pt}{0.400pt}}
\multiput(522.65,137.59)(-0.581,0.482){9}{\rule{0.567pt}{0.116pt}}
\multiput(523.82,136.17)(-5.824,6.000){2}{\rule{0.283pt}{0.400pt}}
\put(512,142.67){\rule{1.445pt}{0.400pt}}
\multiput(515.00,142.17)(-3.000,1.000){2}{\rule{0.723pt}{0.400pt}}
\multiput(508.68,142.94)(-0.920,-0.468){5}{\rule{0.800pt}{0.113pt}}
\multiput(510.34,143.17)(-5.340,-4.000){2}{\rule{0.400pt}{0.400pt}}
\multiput(503.93,140.00)(-0.485,7.052){11}{\rule{0.117pt}{5.414pt}}
\multiput(504.17,140.00)(-7.000,81.762){2}{\rule{0.400pt}{2.707pt}}
\multiput(496.93,233.00)(-0.482,0.852){9}{\rule{0.116pt}{0.767pt}}
\multiput(497.17,233.00)(-6.000,8.409){2}{\rule{0.400pt}{0.383pt}}
\multiput(489.92,243.59)(-0.492,0.485){11}{\rule{0.500pt}{0.117pt}}
\multiput(490.96,242.17)(-5.962,7.000){2}{\rule{0.250pt}{0.400pt}}
\multiput(482.92,250.59)(-0.492,0.485){11}{\rule{0.500pt}{0.117pt}}
\multiput(483.96,249.17)(-5.962,7.000){2}{\rule{0.250pt}{0.400pt}}
\multiput(475.92,257.59)(-0.491,0.482){9}{\rule{0.500pt}{0.116pt}}
\multiput(476.96,256.17)(-4.962,6.000){2}{\rule{0.250pt}{0.400pt}}
\multiput(469.26,263.59)(-0.710,0.477){7}{\rule{0.660pt}{0.115pt}}
\multiput(470.63,262.17)(-5.630,5.000){2}{\rule{0.330pt}{0.400pt}}
\multiput(462.26,268.59)(-0.710,0.477){7}{\rule{0.660pt}{0.115pt}}
\multiput(463.63,267.17)(-5.630,5.000){2}{\rule{0.330pt}{0.400pt}}
\multiput(455.09,273.60)(-0.774,0.468){5}{\rule{0.700pt}{0.113pt}}
\multiput(456.55,272.17)(-4.547,4.000){2}{\rule{0.350pt}{0.400pt}}
\multiput(448.68,277.60)(-0.920,0.468){5}{\rule{0.800pt}{0.113pt}}
\multiput(450.34,276.17)(-5.340,4.000){2}{\rule{0.400pt}{0.400pt}}
\multiput(441.68,281.60)(-0.920,0.468){5}{\rule{0.800pt}{0.113pt}}
\multiput(443.34,280.17)(-5.340,4.000){2}{\rule{0.400pt}{0.400pt}}
\multiput(435.09,285.60)(-0.774,0.468){5}{\rule{0.700pt}{0.113pt}}
\multiput(436.55,284.17)(-4.547,4.000){2}{\rule{0.350pt}{0.400pt}}
\multiput(428.68,289.60)(-0.920,0.468){5}{\rule{0.800pt}{0.113pt}}
\multiput(430.34,288.17)(-5.340,4.000){2}{\rule{0.400pt}{0.400pt}}
\multiput(421.68,293.60)(-0.920,0.468){5}{\rule{0.800pt}{0.113pt}}
\multiput(423.34,292.17)(-5.340,4.000){2}{\rule{0.400pt}{0.400pt}}
\multiput(414.68,297.60)(-0.920,0.468){5}{\rule{0.800pt}{0.113pt}}
\multiput(416.34,296.17)(-5.340,4.000){2}{\rule{0.400pt}{0.400pt}}
\multiput(408.09,301.60)(-0.774,0.468){5}{\rule{0.700pt}{0.113pt}}
\multiput(409.55,300.17)(-4.547,4.000){2}{\rule{0.350pt}{0.400pt}}
\multiput(401.68,305.60)(-0.920,0.468){5}{\rule{0.800pt}{0.113pt}}
\multiput(403.34,304.17)(-5.340,4.000){2}{\rule{0.400pt}{0.400pt}}
\multiput(395.26,309.59)(-0.710,0.477){7}{\rule{0.660pt}{0.115pt}}
\multiput(396.63,308.17)(-5.630,5.000){2}{\rule{0.330pt}{0.400pt}}
\multiput(388.09,314.60)(-0.774,0.468){5}{\rule{0.700pt}{0.113pt}}
\multiput(389.55,313.17)(-4.547,4.000){2}{\rule{0.350pt}{0.400pt}}
\multiput(382.26,318.59)(-0.710,0.477){7}{\rule{0.660pt}{0.115pt}}
\multiput(383.63,317.17)(-5.630,5.000){2}{\rule{0.330pt}{0.400pt}}
\multiput(375.65,323.59)(-0.581,0.482){9}{\rule{0.567pt}{0.116pt}}
\multiput(376.82,322.17)(-5.824,6.000){2}{\rule{0.283pt}{0.400pt}}
\multiput(368.59,329.59)(-0.599,0.477){7}{\rule{0.580pt}{0.115pt}}
\multiput(369.80,328.17)(-4.796,5.000){2}{\rule{0.290pt}{0.400pt}}
\multiput(362.65,334.59)(-0.581,0.482){9}{\rule{0.567pt}{0.116pt}}
\multiput(363.82,333.17)(-5.824,6.000){2}{\rule{0.283pt}{0.400pt}}
\multiput(355.92,340.59)(-0.492,0.485){11}{\rule{0.500pt}{0.117pt}}
\multiput(356.96,339.17)(-5.962,7.000){2}{\rule{0.250pt}{0.400pt}}
\multiput(349.93,347.00)(-0.482,0.581){9}{\rule{0.116pt}{0.567pt}}
\multiput(350.17,347.00)(-6.000,5.824){2}{\rule{0.400pt}{0.283pt}}
\multiput(342.92,354.59)(-0.492,0.485){11}{\rule{0.500pt}{0.117pt}}
\multiput(343.96,353.17)(-5.962,7.000){2}{\rule{0.250pt}{0.400pt}}
\multiput(336.93,361.00)(-0.485,0.569){11}{\rule{0.117pt}{0.557pt}}
\multiput(337.17,361.00)(-7.000,6.844){2}{\rule{0.400pt}{0.279pt}}
\multiput(329.93,369.00)(-0.482,0.581){9}{\rule{0.116pt}{0.567pt}}
\multiput(330.17,369.00)(-6.000,5.824){2}{\rule{0.400pt}{0.283pt}}
\multiput(322.92,376.59)(-0.492,0.485){11}{\rule{0.500pt}{0.117pt}}
\multiput(323.96,375.17)(-5.962,7.000){2}{\rule{0.250pt}{0.400pt}}
\multiput(315.65,383.59)(-0.581,0.482){9}{\rule{0.567pt}{0.116pt}}
\multiput(316.82,382.17)(-5.824,6.000){2}{\rule{0.283pt}{0.400pt}}
\multiput(308.92,389.59)(-0.491,0.482){9}{\rule{0.500pt}{0.116pt}}
\multiput(309.96,388.17)(-4.962,6.000){2}{\rule{0.250pt}{0.400pt}}
\multiput(301.68,395.60)(-0.920,0.468){5}{\rule{0.800pt}{0.113pt}}
\multiput(303.34,394.17)(-5.340,4.000){2}{\rule{0.400pt}{0.400pt}}
\multiput(294.68,399.60)(-0.920,0.468){5}{\rule{0.800pt}{0.113pt}}
\multiput(296.34,398.17)(-5.340,4.000){2}{\rule{0.400pt}{0.400pt}}
\put(285,403.17){\rule{1.300pt}{0.400pt}}
\multiput(288.30,402.17)(-3.302,2.000){2}{\rule{0.650pt}{0.400pt}}
\put(278,405.17){\rule{1.500pt}{0.400pt}}
\multiput(281.89,404.17)(-3.887,2.000){2}{\rule{0.750pt}{0.400pt}}
\put(271,406.67){\rule{1.686pt}{0.400pt}}
\multiput(274.50,406.17)(-3.500,1.000){2}{\rule{0.843pt}{0.400pt}}
\put(265,407.67){\rule{1.445pt}{0.400pt}}
\multiput(268.00,407.17)(-3.000,1.000){2}{\rule{0.723pt}{0.400pt}}
\put(258,408.67){\rule{1.686pt}{0.400pt}}
\multiput(261.50,408.17)(-3.500,1.000){2}{\rule{0.843pt}{0.400pt}}
\put(639,131){\makebox(0,0){$\bullet$}}
\put(632,131){\makebox(0,0){$\bullet$}}
\put(625,131){\makebox(0,0){$\bullet$}}
\put(619,131){\makebox(0,0){$\bullet$}}
\put(612,131){\makebox(0,0){$\bullet$}}
\put(605,131){\makebox(0,0){$\bullet$}}
\put(599,131){\makebox(0,0){$\bullet$}}
\put(592,131){\makebox(0,0){$\bullet$}}
\put(585,131){\makebox(0,0){$\bullet$}}
\put(578,131){\makebox(0,0){$\bullet$}}
\put(572,131){\makebox(0,0){$\bullet$}}
\put(565,131){\makebox(0,0){$\bullet$}}
\put(558,131){\makebox(0,0){$\bullet$}}
\put(552,132){\makebox(0,0){$\bullet$}}
\put(545,132){\makebox(0,0){$\bullet$}}
\put(538,133){\makebox(0,0){$\bullet$}}
\put(532,132){\makebox(0,0){$\bullet$}}
\put(525,137){\makebox(0,0){$\bullet$}}
\put(518,143){\makebox(0,0){$\bullet$}}
\put(512,144){\makebox(0,0){$\bullet$}}
\put(505,140){\makebox(0,0){$\bullet$}}
\put(498,233){\makebox(0,0){$\bullet$}}
\put(492,243){\makebox(0,0){$\bullet$}}
\put(485,250){\makebox(0,0){$\bullet$}}
\put(478,257){\makebox(0,0){$\bullet$}}
\put(472,263){\makebox(0,0){$\bullet$}}
\put(465,268){\makebox(0,0){$\bullet$}}
\put(458,273){\makebox(0,0){$\bullet$}}
\put(452,277){\makebox(0,0){$\bullet$}}
\put(445,281){\makebox(0,0){$\bullet$}}
\put(438,285){\makebox(0,0){$\bullet$}}
\put(432,289){\makebox(0,0){$\bullet$}}
\put(425,293){\makebox(0,0){$\bullet$}}
\put(418,297){\makebox(0,0){$\bullet$}}
\put(411,301){\makebox(0,0){$\bullet$}}
\put(405,305){\makebox(0,0){$\bullet$}}
\put(398,309){\makebox(0,0){$\bullet$}}
\put(391,314){\makebox(0,0){$\bullet$}}
\put(385,318){\makebox(0,0){$\bullet$}}
\put(378,323){\makebox(0,0){$\bullet$}}
\put(371,329){\makebox(0,0){$\bullet$}}
\put(365,334){\makebox(0,0){$\bullet$}}
\put(358,340){\makebox(0,0){$\bullet$}}
\put(351,347){\makebox(0,0){$\bullet$}}
\put(345,354){\makebox(0,0){$\bullet$}}
\put(338,361){\makebox(0,0){$\bullet$}}
\put(331,369){\makebox(0,0){$\bullet$}}
\put(325,376){\makebox(0,0){$\bullet$}}
\put(318,383){\makebox(0,0){$\bullet$}}
\put(311,389){\makebox(0,0){$\bullet$}}
\put(305,395){\makebox(0,0){$\bullet$}}
\put(298,399){\makebox(0,0){$\bullet$}}
\put(291,403){\makebox(0,0){$\bullet$}}
\put(285,405){\makebox(0,0){$\bullet$}}
\put(278,407){\makebox(0,0){$\bullet$}}
\put(271,408){\makebox(0,0){$\bullet$}}
\put(265,409){\makebox(0,0){$\bullet$}}
\put(258,410){\makebox(0,0){$\bullet$}}
\put(545.0,132.0){\rule[-0.200pt]{1.686pt}{0.400pt}}
\put(572,132){\usebox{\plotpoint}}
\put(572.00,132.00){\usebox{\plotpoint}}
\put(551.32,131.00){\usebox{\plotpoint}}
\put(530.64,132.00){\usebox{\plotpoint}}
\put(510.05,131.56){\usebox{\plotpoint}}
\put(490.77,138.88){\usebox{\plotpoint}}
\multiput(478,148)(-4.503,20.261){2}{\usebox{\plotpoint}}
\multiput(472,175)(-2.769,20.570){2}{\usebox{\plotpoint}}
\put(459.76,232.99){\usebox{\plotpoint}}
\put(444.19,246.58){\usebox{\plotpoint}}
\put(426.54,257.34){\usebox{\plotpoint}}
\put(407.96,266.52){\usebox{\plotpoint}}
\put(389.41,275.79){\usebox{\plotpoint}}
\put(370.88,285.08){\usebox{\plotpoint}}
\put(353.12,295.79){\usebox{\plotpoint}}
\put(337.13,308.87){\usebox{\plotpoint}}
\put(323.24,324.26){\usebox{\plotpoint}}
\put(310.99,341.01){\usebox{\plotpoint}}
\put(300.85,359.12){\usebox{\plotpoint}}
\put(290.42,377.06){\usebox{\plotpoint}}
\put(278.79,394.10){\usebox{\plotpoint}}
\put(262.64,406.67){\usebox{\plotpoint}}
\put(258,408){\usebox{\plotpoint}}
\put(572,132){\makebox(0,0){$\circ$}}
\put(565,131){\makebox(0,0){$\circ$}}
\put(558,131){\makebox(0,0){$\circ$}}
\put(552,131){\makebox(0,0){$\circ$}}
\put(545,131){\makebox(0,0){$\circ$}}
\put(538,131){\makebox(0,0){$\circ$}}
\put(532,132){\makebox(0,0){$\circ$}}
\put(525,132){\makebox(0,0){$\circ$}}
\put(518,132){\makebox(0,0){$\circ$}}
\put(512,131){\makebox(0,0){$\circ$}}
\put(505,133){\makebox(0,0){$\circ$}}
\put(498,135){\makebox(0,0){$\circ$}}
\put(492,138){\makebox(0,0){$\circ$}}
\put(485,143){\makebox(0,0){$\circ$}}
\put(478,148){\makebox(0,0){$\circ$}}
\put(472,175){\makebox(0,0){$\circ$}}
\put(465,227){\makebox(0,0){$\circ$}}
\put(458,235){\makebox(0,0){$\circ$}}
\put(452,241){\makebox(0,0){$\circ$}}
\put(445,246){\makebox(0,0){$\circ$}}
\put(438,251){\makebox(0,0){$\circ$}}
\put(432,255){\makebox(0,0){$\circ$}}
\put(425,258){\makebox(0,0){$\circ$}}
\put(418,262){\makebox(0,0){$\circ$}}
\put(411,265){\makebox(0,0){$\circ$}}
\put(405,268){\makebox(0,0){$\circ$}}
\put(398,271){\makebox(0,0){$\circ$}}
\put(391,275){\makebox(0,0){$\circ$}}
\put(385,278){\makebox(0,0){$\circ$}}
\put(378,281){\makebox(0,0){$\circ$}}
\put(371,285){\makebox(0,0){$\circ$}}
\put(365,289){\makebox(0,0){$\circ$}}
\put(358,293){\makebox(0,0){$\circ$}}
\put(351,297){\makebox(0,0){$\circ$}}
\put(345,303){\makebox(0,0){$\circ$}}
\put(338,308){\makebox(0,0){$\circ$}}
\put(331,315){\makebox(0,0){$\circ$}}
\put(325,322){\makebox(0,0){$\circ$}}
\put(318,331){\makebox(0,0){$\circ$}}
\put(311,341){\makebox(0,0){$\circ$}}
\put(305,352){\makebox(0,0){$\circ$}}
\put(298,364){\makebox(0,0){$\circ$}}
\put(291,376){\makebox(0,0){$\circ$}}
\put(285,387){\makebox(0,0){$\circ$}}
\put(278,395){\makebox(0,0){$\circ$}}
\put(271,402){\makebox(0,0){$\circ$}}
\put(265,406){\makebox(0,0){$\circ$}}
\put(258,408){\makebox(0,0){$\circ$}}
\put(171.0,131.0){\rule[-0.200pt]{0.400pt}{67.211pt}}
\put(171.0,131.0){\rule[-0.200pt]{160.921pt}{0.400pt}}
\put(839.0,131.0){\rule[-0.200pt]{0.400pt}{67.211pt}}
\put(171.0,410.0){\rule[-0.200pt]{160.921pt}{0.400pt}}
\end{picture}

\begin{picture}(900,450)(0,0)
\put(171.0,131.0){\rule[-0.200pt]{4.818pt}{0.400pt}}
\put(151,131){\makebox(0,0)[r]{-4}}
\put(819.0,131.0){\rule[-0.200pt]{4.818pt}{0.400pt}}
\put(171.0,162.0){\rule[-0.200pt]{4.818pt}{0.400pt}}
%\put(151,162){\makebox(0,0)[r]{-3.5}}
\put(819.0,162.0){\rule[-0.200pt]{4.818pt}{0.400pt}}
\put(171.0,193.0){\rule[-0.200pt]{4.818pt}{0.400pt}}
\put(151,193){\makebox(0,0)[r]{-3}}
\put(819.0,193.0){\rule[-0.200pt]{4.818pt}{0.400pt}}
\put(171.0,224.0){\rule[-0.200pt]{4.818pt}{0.400pt}}
%\put(151,224){\makebox(0,0)[r]{-2.5}}
\put(819.0,224.0){\rule[-0.200pt]{4.818pt}{0.400pt}}
\put(171.0,255.0){\rule[-0.200pt]{4.818pt}{0.400pt}}
\put(151,255){\makebox(0,0)[r]{-2}}
\put(819.0,255.0){\rule[-0.200pt]{4.818pt}{0.400pt}}
\put(171.0,286.0){\rule[-0.200pt]{4.818pt}{0.400pt}}
%\put(151,286){\makebox(0,0)[r]{-1.5}}
\put(819.0,286.0){\rule[-0.200pt]{4.818pt}{0.400pt}}
\put(171.0,317.0){\rule[-0.200pt]{4.818pt}{0.400pt}}
\put(151,317){\makebox(0,0)[r]{-1}}
\put(819.0,317.0){\rule[-0.200pt]{4.818pt}{0.400pt}}
\put(171.0,348.0){\rule[-0.200pt]{4.818pt}{0.400pt}}
%\put(151,348){\makebox(0,0)[r]{-0.5}}
\put(819.0,348.0){\rule[-0.200pt]{4.818pt}{0.400pt}}
\put(171.0,379.0){\rule[-0.200pt]{4.818pt}{0.400pt}}
\put(151,379){\makebox(0,0)[r]{ 0}}
\put(819.0,379.0){\rule[-0.200pt]{4.818pt}{0.400pt}}
\put(171.0,410.0){\rule[-0.200pt]{4.818pt}{0.400pt}}
%\put(151,410){\makebox(0,0)[r]{ 0.5}}
\put(819.0,410.0){\rule[-0.200pt]{4.818pt}{0.400pt}}
\put(171.0,131.0){\rule[-0.200pt]{0.400pt}{4.818pt}}
\put(171,90){\makebox(0,0){ 0}}
\put(171.0,390.0){\rule[-0.200pt]{0.400pt}{4.818pt}}
\put(305.0,131.0){\rule[-0.200pt]{0.400pt}{4.818pt}}
\put(305,90){\makebox(0,0){ 1}}
\put(305.0,390.0){\rule[-0.200pt]{0.400pt}{4.818pt}}
\put(438.0,131.0){\rule[-0.200pt]{0.400pt}{4.818pt}}
\put(438,90){\makebox(0,0){ 2}}
\put(438.0,390.0){\rule[-0.200pt]{0.400pt}{4.818pt}}
\put(572.0,131.0){\rule[-0.200pt]{0.400pt}{4.818pt}}
\put(572,90){\makebox(0,0){ 3}}
\put(572.0,390.0){\rule[-0.200pt]{0.400pt}{4.818pt}}
\put(705.0,131.0){\rule[-0.200pt]{0.400pt}{4.818pt}}
\put(705,90){\makebox(0,0){ 4}}
\put(705.0,390.0){\rule[-0.200pt]{0.400pt}{4.818pt}}
\put(839.0,131.0){\rule[-0.200pt]{0.400pt}{4.818pt}}
\put(839,90){\makebox(0,0){ 5}}
\put(839.0,390.0){\rule[-0.200pt]{0.400pt}{4.818pt}}
\put(171.0,131.0){\rule[-0.200pt]{0.400pt}{67.211pt}}
\put(171.0,131.0){\rule[-0.200pt]{160.921pt}{0.400pt}}
\put(839.0,131.0){\rule[-0.200pt]{0.400pt}{67.211pt}}
\put(171.0,410.0){\rule[-0.200pt]{160.921pt}{0.400pt}}
\put(30,270){\makebox(0,0){${{dQ_s} \over {dT}}$}}
\put(220,350){\makebox(0,0){(c)}}
\put(505,29){\makebox(0,0){$T$}}
\put(632,379){\usebox{\plotpoint}}
\put(599,378.67){\rule{1.445pt}{0.400pt}}
\multiput(602.00,378.17)(-3.000,1.000){2}{\rule{0.723pt}{0.400pt}}
\put(592,378.67){\rule{1.686pt}{0.400pt}}
\multiput(595.50,379.17)(-3.500,-1.000){2}{\rule{0.843pt}{0.400pt}}
\put(605.0,379.0){\rule[-0.200pt]{6.504pt}{0.400pt}}
\put(558,377.17){\rule{1.500pt}{0.400pt}}
\multiput(561.89,378.17)(-3.887,-2.000){2}{\rule{0.750pt}{0.400pt}}
\put(565.0,379.0){\rule[-0.200pt]{6.504pt}{0.400pt}}
\put(545,376.67){\rule{1.686pt}{0.400pt}}
\multiput(548.50,376.17)(-3.500,1.000){2}{\rule{0.843pt}{0.400pt}}
\put(552.0,377.0){\rule[-0.200pt]{1.445pt}{0.400pt}}
\multiput(536.93,375.37)(-0.482,-0.671){9}{\rule{0.116pt}{0.633pt}}
\multiput(537.17,376.69)(-6.000,-6.685){2}{\rule{0.400pt}{0.317pt}}
\multiput(530.93,366.03)(-0.485,-1.103){11}{\rule{0.117pt}{0.957pt}}
\multiput(531.17,368.01)(-7.000,-13.013){2}{\rule{0.400pt}{0.479pt}}
\multiput(522.92,355.59)(-0.492,0.485){11}{\rule{0.500pt}{0.117pt}}
\multiput(523.96,354.17)(-5.962,7.000){2}{\rule{0.250pt}{0.400pt}}
\multiput(516.93,362.00)(-0.482,2.208){9}{\rule{0.116pt}{1.767pt}}
\multiput(517.17,362.00)(-6.000,21.333){2}{\rule{0.400pt}{0.883pt}}
\multiput(510.93,337.72)(-0.485,-15.672){11}{\rule{0.117pt}{11.871pt}}
\multiput(511.17,362.36)(-7.000,-181.360){2}{\rule{0.400pt}{5.936pt}}
\multiput(503.93,173.23)(-0.485,-2.323){11}{\rule{0.117pt}{1.871pt}}
\multiput(504.17,177.12)(-7.000,-27.116){2}{\rule{0.400pt}{0.936pt}}
\multiput(496.93,150.00)(-0.482,17.306){9}{\rule{0.116pt}{12.900pt}}
\multiput(497.17,150.00)(-6.000,165.225){2}{\rule{0.400pt}{6.450pt}}
\multiput(489.65,342.59)(-0.581,0.482){9}{\rule{0.567pt}{0.116pt}}
\multiput(490.82,341.17)(-5.824,6.000){2}{\rule{0.283pt}{0.400pt}}
\multiput(480.71,348.61)(-1.355,0.447){3}{\rule{1.033pt}{0.108pt}}
\multiput(482.86,347.17)(-4.855,3.000){2}{\rule{0.517pt}{0.400pt}}
\multiput(474.26,351.61)(-1.132,0.447){3}{\rule{0.900pt}{0.108pt}}
\multiput(476.13,350.17)(-4.132,3.000){2}{\rule{0.450pt}{0.400pt}}
\multiput(467.71,354.61)(-1.355,0.447){3}{\rule{1.033pt}{0.108pt}}
\multiput(469.86,353.17)(-4.855,3.000){2}{\rule{0.517pt}{0.400pt}}
\put(458,356.67){\rule{1.686pt}{0.400pt}}
\multiput(461.50,356.17)(-3.500,1.000){2}{\rule{0.843pt}{0.400pt}}
\put(452,358.17){\rule{1.300pt}{0.400pt}}
\multiput(455.30,357.17)(-3.302,2.000){2}{\rule{0.650pt}{0.400pt}}
\put(445,359.67){\rule{1.686pt}{0.400pt}}
\multiput(448.50,359.17)(-3.500,1.000){2}{\rule{0.843pt}{0.400pt}}
\put(538.0,378.0){\rule[-0.200pt]{1.686pt}{0.400pt}}
\put(425,360.67){\rule{1.686pt}{0.400pt}}
\multiput(428.50,360.17)(-3.500,1.000){2}{\rule{0.843pt}{0.400pt}}
\put(418,360.67){\rule{1.686pt}{0.400pt}}
\multiput(421.50,361.17)(-3.500,-1.000){2}{\rule{0.843pt}{0.400pt}}
\put(432.0,361.0){\rule[-0.200pt]{3.132pt}{0.400pt}}
\put(398,359.67){\rule{1.686pt}{0.400pt}}
\multiput(401.50,360.17)(-3.500,-1.000){2}{\rule{0.843pt}{0.400pt}}
\put(391,358.67){\rule{1.686pt}{0.400pt}}
\multiput(394.50,359.17)(-3.500,-1.000){2}{\rule{0.843pt}{0.400pt}}
\put(385,357.67){\rule{1.445pt}{0.400pt}}
\multiput(388.00,358.17)(-3.000,-1.000){2}{\rule{0.723pt}{0.400pt}}
\put(378,356.17){\rule{1.500pt}{0.400pt}}
\multiput(381.89,357.17)(-3.887,-2.000){2}{\rule{0.750pt}{0.400pt}}
\put(371,354.67){\rule{1.686pt}{0.400pt}}
\multiput(374.50,355.17)(-3.500,-1.000){2}{\rule{0.843pt}{0.400pt}}
\put(365,353.17){\rule{1.300pt}{0.400pt}}
\multiput(368.30,354.17)(-3.302,-2.000){2}{\rule{0.650pt}{0.400pt}}
\put(358,351.17){\rule{1.500pt}{0.400pt}}
\multiput(361.89,352.17)(-3.887,-2.000){2}{\rule{0.750pt}{0.400pt}}
\put(351,349.17){\rule{1.500pt}{0.400pt}}
\multiput(354.89,350.17)(-3.887,-2.000){2}{\rule{0.750pt}{0.400pt}}
\put(345,347.17){\rule{1.300pt}{0.400pt}}
\multiput(348.30,348.17)(-3.302,-2.000){2}{\rule{0.650pt}{0.400pt}}
\put(338,345.67){\rule{1.686pt}{0.400pt}}
\multiput(341.50,346.17)(-3.500,-1.000){2}{\rule{0.843pt}{0.400pt}}
\put(405.0,361.0){\rule[-0.200pt]{3.132pt}{0.400pt}}
\put(325,345.67){\rule{1.445pt}{0.400pt}}
\multiput(328.00,345.17)(-3.000,1.000){2}{\rule{0.723pt}{0.400pt}}
\put(318,347.17){\rule{1.500pt}{0.400pt}}
\multiput(321.89,346.17)(-3.887,2.000){2}{\rule{0.750pt}{0.400pt}}
\multiput(313.71,349.61)(-1.355,0.447){3}{\rule{1.033pt}{0.108pt}}
\multiput(315.86,348.17)(-4.855,3.000){2}{\rule{0.517pt}{0.400pt}}
\multiput(308.09,352.60)(-0.774,0.468){5}{\rule{0.700pt}{0.113pt}}
\multiput(309.55,351.17)(-4.547,4.000){2}{\rule{0.350pt}{0.400pt}}
\multiput(302.26,356.59)(-0.710,0.477){7}{\rule{0.660pt}{0.115pt}}
\multiput(303.63,355.17)(-5.630,5.000){2}{\rule{0.330pt}{0.400pt}}
\multiput(295.26,361.59)(-0.710,0.477){7}{\rule{0.660pt}{0.115pt}}
\multiput(296.63,360.17)(-5.630,5.000){2}{\rule{0.330pt}{0.400pt}}
\multiput(288.09,366.60)(-0.774,0.468){5}{\rule{0.700pt}{0.113pt}}
\multiput(289.55,365.17)(-4.547,4.000){2}{\rule{0.350pt}{0.400pt}}
\multiput(280.71,370.61)(-1.355,0.447){3}{\rule{1.033pt}{0.108pt}}
\multiput(282.86,369.17)(-4.855,3.000){2}{\rule{0.517pt}{0.400pt}}
\put(271,373.17){\rule{1.500pt}{0.400pt}}
\multiput(274.89,372.17)(-3.887,2.000){2}{\rule{0.750pt}{0.400pt}}
\put(265,374.67){\rule{1.445pt}{0.400pt}}
\multiput(268.00,374.17)(-3.000,1.000){2}{\rule{0.723pt}{0.400pt}}
\put(632,379){\makebox(0,0){$\bullet$}}
\put(625,379){\makebox(0,0){$\bullet$}}
\put(619,379){\makebox(0,0){$\bullet$}}
\put(612,379){\makebox(0,0){$\bullet$}}
\put(605,379){\makebox(0,0){$\bullet$}}
\put(599,380){\makebox(0,0){$\bullet$}}
\put(592,379){\makebox(0,0){$\bullet$}}
\put(585,379){\makebox(0,0){$\bullet$}}
\put(578,379){\makebox(0,0){$\bullet$}}
\put(572,379){\makebox(0,0){$\bullet$}}
\put(565,379){\makebox(0,0){$\bullet$}}
\put(558,377){\makebox(0,0){$\bullet$}}
\put(552,377){\makebox(0,0){$\bullet$}}
\put(545,378){\makebox(0,0){$\bullet$}}
\put(538,378){\makebox(0,0){$\bullet$}}
\put(532,370){\makebox(0,0){$\bullet$}}
\put(525,355){\makebox(0,0){$\bullet$}}
\put(518,362){\makebox(0,0){$\bullet$}}
\put(512,387){\makebox(0,0){$\bullet$}}
\put(505,181){\makebox(0,0){$\bullet$}}
\put(498,150){\makebox(0,0){$\bullet$}}
\put(492,342){\makebox(0,0){$\bullet$}}
\put(485,348){\makebox(0,0){$\bullet$}}
\put(478,351){\makebox(0,0){$\bullet$}}
\put(472,354){\makebox(0,0){$\bullet$}}
\put(465,357){\makebox(0,0){$\bullet$}}
\put(458,358){\makebox(0,0){$\bullet$}}
\put(452,360){\makebox(0,0){$\bullet$}}
\put(445,361){\makebox(0,0){$\bullet$}}
\put(438,361){\makebox(0,0){$\bullet$}}
\put(432,361){\makebox(0,0){$\bullet$}}
\put(425,362){\makebox(0,0){$\bullet$}}
\put(418,361){\makebox(0,0){$\bullet$}}
\put(411,361){\makebox(0,0){$\bullet$}}
\put(405,361){\makebox(0,0){$\bullet$}}
\put(398,360){\makebox(0,0){$\bullet$}}
\put(391,359){\makebox(0,0){$\bullet$}}
\put(385,358){\makebox(0,0){$\bullet$}}
\put(378,356){\makebox(0,0){$\bullet$}}
\put(371,355){\makebox(0,0){$\bullet$}}
\put(365,353){\makebox(0,0){$\bullet$}}
\put(358,351){\makebox(0,0){$\bullet$}}
\put(351,349){\makebox(0,0){$\bullet$}}
\put(345,347){\makebox(0,0){$\bullet$}}
\put(338,346){\makebox(0,0){$\bullet$}}
\put(331,346){\makebox(0,0){$\bullet$}}
\put(325,347){\makebox(0,0){$\bullet$}}
\put(318,349){\makebox(0,0){$\bullet$}}
\put(311,352){\makebox(0,0){$\bullet$}}
\put(305,356){\makebox(0,0){$\bullet$}}
\put(298,361){\makebox(0,0){$\bullet$}}
\put(291,366){\makebox(0,0){$\bullet$}}
\put(285,370){\makebox(0,0){$\bullet$}}
\put(278,373){\makebox(0,0){$\bullet$}}
\put(271,375){\makebox(0,0){$\bullet$}}
\put(265,376){\makebox(0,0){$\bullet$}}
\put(331.0,346.0){\rule[-0.200pt]{1.686pt}{0.400pt}}
\put(565,381){\usebox{\plotpoint}}
\put(565.00,381.00){\usebox{\plotpoint}}
\put(544.54,378.87){\usebox{\plotpoint}}
\put(524.42,380.08){\usebox{\plotpoint}}
\put(506.64,372.40){\usebox{\plotpoint}}
\put(490.79,359.48){\usebox{\plotpoint}}
\multiput(485,357)(-2.995,-20.538){2}{\usebox{\plotpoint}}
\multiput(478,309)(-1.162,-20.723){5}{\usebox{\plotpoint}}
\multiput(472,202)(-3.335,20.486){2}{\usebox{\plotpoint}}
\multiput(465,245)(-1.407,20.708){5}{\usebox{\plotpoint}}
\put(456.19,349.81){\usebox{\plotpoint}}
\put(438.79,360.66){\usebox{\plotpoint}}
\put(418.25,362.96){\usebox{\plotpoint}}
\put(398.38,364.11){\usebox{\plotpoint}}
\put(378.00,363.00){\usebox{\plotpoint}}
\put(357.51,359.79){\usebox{\plotpoint}}
\put(338.41,352.24){\usebox{\plotpoint}}
\put(321.61,340.09){\usebox{\plotpoint}}
\put(304.81,327.97){\usebox{\plotpoint}}
\put(286.57,333.17){\usebox{\plotpoint}}
\put(274.96,350.34){\usebox{\plotpoint}}
\put(265,366){\usebox{\plotpoint}}
\put(565,381){\makebox(0,0){$\circ$}}
\put(558,379){\makebox(0,0){$\circ$}}
\put(552,379){\makebox(0,0){$\circ$}}
\put(545,379){\makebox(0,0){$\circ$}}
\put(538,377){\makebox(0,0){$\circ$}}
\put(532,378){\makebox(0,0){$\circ$}}
\put(525,380){\makebox(0,0){$\circ$}}
\put(518,381){\makebox(0,0){$\circ$}}
\put(512,377){\makebox(0,0){$\circ$}}
\put(505,371){\makebox(0,0){$\circ$}}
\put(498,367){\makebox(0,0){$\circ$}}
\put(492,360){\makebox(0,0){$\circ$}}
\put(485,357){\makebox(0,0){$\circ$}}
\put(478,309){\makebox(0,0){$\circ$}}
\put(472,202){\makebox(0,0){$\circ$}}
\put(465,245){\makebox(0,0){$\circ$}}
\put(458,348){\makebox(0,0){$\circ$}}
\put(452,354){\makebox(0,0){$\circ$}}
\put(445,358){\makebox(0,0){$\circ$}}
\put(438,361){\makebox(0,0){$\circ$}}
\put(432,362){\makebox(0,0){$\circ$}}
\put(425,362){\makebox(0,0){$\circ$}}
\put(418,363){\makebox(0,0){$\circ$}}
\put(411,366){\makebox(0,0){$\circ$}}
\put(405,366){\makebox(0,0){$\circ$}}
\put(398,364){\makebox(0,0){$\circ$}}
\put(391,364){\makebox(0,0){$\circ$}}
\put(385,365){\makebox(0,0){$\circ$}}
\put(378,363){\makebox(0,0){$\circ$}}
\put(371,362){\makebox(0,0){$\circ$}}
\put(365,361){\makebox(0,0){$\circ$}}
\put(358,360){\makebox(0,0){$\circ$}}
\put(351,357){\makebox(0,0){$\circ$}}
\put(345,356){\makebox(0,0){$\circ$}}
\put(338,352){\makebox(0,0){$\circ$}}
\put(331,347){\makebox(0,0){$\circ$}}
\put(325,343){\makebox(0,0){$\circ$}}
\put(318,337){\makebox(0,0){$\circ$}}
\put(311,332){\makebox(0,0){$\circ$}}
\put(305,328){\makebox(0,0){$\circ$}}
\put(298,327){\makebox(0,0){$\circ$}}
\put(291,328){\makebox(0,0){$\circ$}}
\put(285,335){\makebox(0,0){$\circ$}}
\put(278,346){\makebox(0,0){$\circ$}}
\put(271,356){\makebox(0,0){$\circ$}}
\put(265,366){\makebox(0,0){$\circ$}}
\put(171.0,131.0){\rule[-0.200pt]{0.400pt}{67.211pt}}
\put(171.0,131.0){\rule[-0.200pt]{160.921pt}{0.400pt}}
\put(839.0,131.0){\rule[-0.200pt]{0.400pt}{67.211pt}}
\put(171.0,410.0){\rule[-0.200pt]{160.921pt}{0.400pt}}
\end{picture}

\begin{picture}(900,450)(0,0)
\put(171.0,131.0){\rule[-0.200pt]{4.818pt}{0.400pt}}
\put(151,131){\makebox(0,0)[r]{-3}}
\put(819.0,131.0){\rule[-0.200pt]{4.818pt}{0.400pt}}
\put(171.0,187.0){\rule[-0.200pt]{4.818pt}{0.400pt}}
\put(151,187){\makebox(0,0)[r]{-2.5}}
\put(819.0,187.0){\rule[-0.200pt]{4.818pt}{0.400pt}}
\put(171.0,243.0){\rule[-0.200pt]{4.818pt}{0.400pt}}
\put(151,243){\makebox(0,0)[r]{-2}}
\put(819.0,243.0){\rule[-0.200pt]{4.818pt}{0.400pt}}
\put(171.0,298.0){\rule[-0.200pt]{4.818pt}{0.400pt}}
\put(151,298){\makebox(0,0)[r]{-1.5}}
\put(819.0,298.0){\rule[-0.200pt]{4.818pt}{0.400pt}}
\put(171.0,354.0){\rule[-0.200pt]{4.818pt}{0.400pt}}
\put(151,354){\makebox(0,0)[r]{-1}}
\put(819.0,354.0){\rule[-0.200pt]{4.818pt}{0.400pt}}
\put(171.0,410.0){\rule[-0.200pt]{4.818pt}{0.400pt}}
\put(151,410){\makebox(0,0)[r]{-0.5}}
\put(819.0,410.0){\rule[-0.200pt]{4.818pt}{0.400pt}}
\put(171.0,131.0){\rule[-0.200pt]{0.400pt}{4.818pt}}
\put(171,90){\makebox(0,0){ 0}}
\put(171.0,390.0){\rule[-0.200pt]{0.400pt}{4.818pt}}
\put(305.0,131.0){\rule[-0.200pt]{0.400pt}{4.818pt}}
\put(305,90){\makebox(0,0){ 1}}
\put(305.0,390.0){\rule[-0.200pt]{0.400pt}{4.818pt}}
\put(438.0,131.0){\rule[-0.200pt]{0.400pt}{4.818pt}}
\put(438,90){\makebox(0,0){ 2}}
\put(438.0,390.0){\rule[-0.200pt]{0.400pt}{4.818pt}}
\put(572.0,131.0){\rule[-0.200pt]{0.400pt}{4.818pt}}
\put(572,90){\makebox(0,0){ 3}}
\put(572.0,390.0){\rule[-0.200pt]{0.400pt}{4.818pt}}
\put(705.0,131.0){\rule[-0.200pt]{0.400pt}{4.818pt}}
\put(705,90){\makebox(0,0){ 4}}
\put(705.0,390.0){\rule[-0.200pt]{0.400pt}{4.818pt}}
\put(839.0,131.0){\rule[-0.200pt]{0.400pt}{4.818pt}}
\put(839,90){\makebox(0,0){ 5}}
\put(839.0,390.0){\rule[-0.200pt]{0.400pt}{4.818pt}}
\put(171.0,131.0){\rule[-0.200pt]{0.400pt}{67.211pt}}
\put(171.0,131.0){\rule[-0.200pt]{160.921pt}{0.400pt}}
\put(839.0,131.0){\rule[-0.200pt]{0.400pt}{67.211pt}}
\put(171.0,410.0){\rule[-0.200pt]{160.921pt}{0.400pt}}
\put(30,270){\makebox(0,0){$E$}}
\put(220,350){\makebox(0,0){(d)}}
\put(505,29){\makebox(0,0){$T$}}
\put(839,375){\usebox{\plotpoint}}
\put(832,373.17){\rule{1.500pt}{0.400pt}}
\multiput(835.89,374.17)(-3.887,-2.000){2}{\rule{0.750pt}{0.400pt}}
\put(826,371.67){\rule{1.445pt}{0.400pt}}
\multiput(829.00,372.17)(-3.000,-1.000){2}{\rule{0.723pt}{0.400pt}}
\put(819,370.17){\rule{1.500pt}{0.400pt}}
\multiput(822.89,371.17)(-3.887,-2.000){2}{\rule{0.750pt}{0.400pt}}
\put(812,368.67){\rule{1.686pt}{0.400pt}}
\multiput(815.50,369.17)(-3.500,-1.000){2}{\rule{0.843pt}{0.400pt}}
\put(806,367.17){\rule{1.300pt}{0.400pt}}
\multiput(809.30,368.17)(-3.302,-2.000){2}{\rule{0.650pt}{0.400pt}}
\put(799,365.17){\rule{1.500pt}{0.400pt}}
\multiput(802.89,366.17)(-3.887,-2.000){2}{\rule{0.750pt}{0.400pt}}
\put(792,363.17){\rule{1.500pt}{0.400pt}}
\multiput(795.89,364.17)(-3.887,-2.000){2}{\rule{0.750pt}{0.400pt}}
\multiput(788.26,361.95)(-1.132,-0.447){3}{\rule{0.900pt}{0.108pt}}
\multiput(790.13,362.17)(-4.132,-3.000){2}{\rule{0.450pt}{0.400pt}}
\multiput(781.71,358.95)(-1.355,-0.447){3}{\rule{1.033pt}{0.108pt}}
\multiput(783.86,359.17)(-4.855,-3.000){2}{\rule{0.517pt}{0.400pt}}
\multiput(774.71,355.95)(-1.355,-0.447){3}{\rule{1.033pt}{0.108pt}}
\multiput(776.86,356.17)(-4.855,-3.000){2}{\rule{0.517pt}{0.400pt}}
\multiput(769.92,352.93)(-0.491,-0.482){9}{\rule{0.500pt}{0.116pt}}
\multiput(770.96,353.17)(-4.962,-6.000){2}{\rule{0.250pt}{0.400pt}}
\multiput(764.93,345.21)(-0.485,-0.721){11}{\rule{0.117pt}{0.671pt}}
\multiput(765.17,346.61)(-7.000,-8.606){2}{\rule{0.400pt}{0.336pt}}
\multiput(757.93,334.50)(-0.485,-0.950){11}{\rule{0.117pt}{0.843pt}}
\multiput(758.17,336.25)(-7.000,-11.251){2}{\rule{0.400pt}{0.421pt}}
\multiput(750.93,321.74)(-0.485,-0.874){11}{\rule{0.117pt}{0.786pt}}
\multiput(751.17,323.37)(-7.000,-10.369){2}{\rule{0.400pt}{0.393pt}}
\multiput(743.93,309.82)(-0.482,-0.852){9}{\rule{0.116pt}{0.767pt}}
\multiput(744.17,311.41)(-6.000,-8.409){2}{\rule{0.400pt}{0.383pt}}
\multiput(737.93,300.21)(-0.485,-0.721){11}{\rule{0.117pt}{0.671pt}}
\multiput(738.17,301.61)(-7.000,-8.606){2}{\rule{0.400pt}{0.336pt}}
\multiput(730.93,290.45)(-0.485,-0.645){11}{\rule{0.117pt}{0.614pt}}
\multiput(731.17,291.73)(-7.000,-7.725){2}{\rule{0.400pt}{0.307pt}}
\multiput(723.93,281.37)(-0.482,-0.671){9}{\rule{0.116pt}{0.633pt}}
\multiput(724.17,282.69)(-6.000,-6.685){2}{\rule{0.400pt}{0.317pt}}
\multiput(717.93,273.69)(-0.485,-0.569){11}{\rule{0.117pt}{0.557pt}}
\multiput(718.17,274.84)(-7.000,-6.844){2}{\rule{0.400pt}{0.279pt}}
\multiput(710.93,265.69)(-0.485,-0.569){11}{\rule{0.117pt}{0.557pt}}
\multiput(711.17,266.84)(-7.000,-6.844){2}{\rule{0.400pt}{0.279pt}}
\multiput(703.93,257.65)(-0.482,-0.581){9}{\rule{0.116pt}{0.567pt}}
\multiput(704.17,258.82)(-6.000,-5.824){2}{\rule{0.400pt}{0.283pt}}
\multiput(696.65,251.93)(-0.581,-0.482){9}{\rule{0.567pt}{0.116pt}}
\multiput(697.82,252.17)(-5.824,-6.000){2}{\rule{0.283pt}{0.400pt}}
\multiput(689.92,245.93)(-0.492,-0.485){11}{\rule{0.500pt}{0.117pt}}
\multiput(690.96,246.17)(-5.962,-7.000){2}{\rule{0.250pt}{0.400pt}}
\multiput(682.92,238.93)(-0.491,-0.482){9}{\rule{0.500pt}{0.116pt}}
\multiput(683.96,239.17)(-4.962,-6.000){2}{\rule{0.250pt}{0.400pt}}
\multiput(676.65,232.93)(-0.581,-0.482){9}{\rule{0.567pt}{0.116pt}}
\multiput(677.82,233.17)(-5.824,-6.000){2}{\rule{0.283pt}{0.400pt}}
\multiput(669.26,226.93)(-0.710,-0.477){7}{\rule{0.660pt}{0.115pt}}
\multiput(670.63,227.17)(-5.630,-5.000){2}{\rule{0.330pt}{0.400pt}}
\multiput(662.92,221.93)(-0.491,-0.482){9}{\rule{0.500pt}{0.116pt}}
\multiput(663.96,222.17)(-4.962,-6.000){2}{\rule{0.250pt}{0.400pt}}
\multiput(656.26,215.93)(-0.710,-0.477){7}{\rule{0.660pt}{0.115pt}}
\multiput(657.63,216.17)(-5.630,-5.000){2}{\rule{0.330pt}{0.400pt}}
\multiput(648.68,210.94)(-0.920,-0.468){5}{\rule{0.800pt}{0.113pt}}
\multiput(650.34,211.17)(-5.340,-4.000){2}{\rule{0.400pt}{0.400pt}}
\multiput(642.59,206.93)(-0.599,-0.477){7}{\rule{0.580pt}{0.115pt}}
\multiput(643.80,207.17)(-4.796,-5.000){2}{\rule{0.290pt}{0.400pt}}
\multiput(636.26,201.93)(-0.710,-0.477){7}{\rule{0.660pt}{0.115pt}}
\multiput(637.63,202.17)(-5.630,-5.000){2}{\rule{0.330pt}{0.400pt}}
\multiput(628.68,196.94)(-0.920,-0.468){5}{\rule{0.800pt}{0.113pt}}
\multiput(630.34,197.17)(-5.340,-4.000){2}{\rule{0.400pt}{0.400pt}}
\multiput(622.09,192.94)(-0.774,-0.468){5}{\rule{0.700pt}{0.113pt}}
\multiput(623.55,193.17)(-4.547,-4.000){2}{\rule{0.350pt}{0.400pt}}
\multiput(615.68,188.94)(-0.920,-0.468){5}{\rule{0.800pt}{0.113pt}}
\multiput(617.34,189.17)(-5.340,-4.000){2}{\rule{0.400pt}{0.400pt}}
\multiput(607.71,184.95)(-1.355,-0.447){3}{\rule{1.033pt}{0.108pt}}
\multiput(609.86,185.17)(-4.855,-3.000){2}{\rule{0.517pt}{0.400pt}}
\multiput(602.09,181.94)(-0.774,-0.468){5}{\rule{0.700pt}{0.113pt}}
\multiput(603.55,182.17)(-4.547,-4.000){2}{\rule{0.350pt}{0.400pt}}
\multiput(594.71,177.95)(-1.355,-0.447){3}{\rule{1.033pt}{0.108pt}}
\multiput(596.86,178.17)(-4.855,-3.000){2}{\rule{0.517pt}{0.400pt}}
\multiput(587.71,174.95)(-1.355,-0.447){3}{\rule{1.033pt}{0.108pt}}
\multiput(589.86,175.17)(-4.855,-3.000){2}{\rule{0.517pt}{0.400pt}}
\multiput(580.71,171.95)(-1.355,-0.447){3}{\rule{1.033pt}{0.108pt}}
\multiput(582.86,172.17)(-4.855,-3.000){2}{\rule{0.517pt}{0.400pt}}
\multiput(574.26,168.95)(-1.132,-0.447){3}{\rule{0.900pt}{0.108pt}}
\multiput(576.13,169.17)(-4.132,-3.000){2}{\rule{0.450pt}{0.400pt}}
\multiput(567.71,165.95)(-1.355,-0.447){3}{\rule{1.033pt}{0.108pt}}
\multiput(569.86,166.17)(-4.855,-3.000){2}{\rule{0.517pt}{0.400pt}}
\put(558,162.17){\rule{1.500pt}{0.400pt}}
\multiput(561.89,163.17)(-3.887,-2.000){2}{\rule{0.750pt}{0.400pt}}
\multiput(554.26,160.95)(-1.132,-0.447){3}{\rule{0.900pt}{0.108pt}}
\multiput(556.13,161.17)(-4.132,-3.000){2}{\rule{0.450pt}{0.400pt}}
\put(545,157.17){\rule{1.500pt}{0.400pt}}
\multiput(548.89,158.17)(-3.887,-2.000){2}{\rule{0.750pt}{0.400pt}}
\put(538,155.17){\rule{1.500pt}{0.400pt}}
\multiput(541.89,156.17)(-3.887,-2.000){2}{\rule{0.750pt}{0.400pt}}
\put(532,153.17){\rule{1.300pt}{0.400pt}}
\multiput(535.30,154.17)(-3.302,-2.000){2}{\rule{0.650pt}{0.400pt}}
\put(525,151.17){\rule{1.500pt}{0.400pt}}
\multiput(528.89,152.17)(-3.887,-2.000){2}{\rule{0.750pt}{0.400pt}}
\put(518,149.17){\rule{1.500pt}{0.400pt}}
\multiput(521.89,150.17)(-3.887,-2.000){2}{\rule{0.750pt}{0.400pt}}
\put(512,147.17){\rule{1.300pt}{0.400pt}}
\multiput(515.30,148.17)(-3.302,-2.000){2}{\rule{0.650pt}{0.400pt}}
\put(505,145.67){\rule{1.686pt}{0.400pt}}
\multiput(508.50,146.17)(-3.500,-1.000){2}{\rule{0.843pt}{0.400pt}}
\put(498,144.17){\rule{1.500pt}{0.400pt}}
\multiput(501.89,145.17)(-3.887,-2.000){2}{\rule{0.750pt}{0.400pt}}
\put(492,142.67){\rule{1.445pt}{0.400pt}}
\multiput(495.00,143.17)(-3.000,-1.000){2}{\rule{0.723pt}{0.400pt}}
\put(485,141.67){\rule{1.686pt}{0.400pt}}
\multiput(488.50,142.17)(-3.500,-1.000){2}{\rule{0.843pt}{0.400pt}}
\put(478,140.67){\rule{1.686pt}{0.400pt}}
\multiput(481.50,141.17)(-3.500,-1.000){2}{\rule{0.843pt}{0.400pt}}
\put(472,139.17){\rule{1.300pt}{0.400pt}}
\multiput(475.30,140.17)(-3.302,-2.000){2}{\rule{0.650pt}{0.400pt}}
\put(839,375){\makebox(0,0){$\bullet$}}
\put(832,373){\makebox(0,0){$\bullet$}}
\put(826,372){\makebox(0,0){$\bullet$}}
\put(819,370){\makebox(0,0){$\bullet$}}
\put(812,369){\makebox(0,0){$\bullet$}}
\put(806,367){\makebox(0,0){$\bullet$}}
\put(799,365){\makebox(0,0){$\bullet$}}
\put(792,363){\makebox(0,0){$\bullet$}}
\put(786,360){\makebox(0,0){$\bullet$}}
\put(779,357){\makebox(0,0){$\bullet$}}
\put(772,354){\makebox(0,0){$\bullet$}}
\put(766,348){\makebox(0,0){$\bullet$}}
\put(759,338){\makebox(0,0){$\bullet$}}
\put(752,325){\makebox(0,0){$\bullet$}}
\put(745,313){\makebox(0,0){$\bullet$}}
\put(739,303){\makebox(0,0){$\bullet$}}
\put(732,293){\makebox(0,0){$\bullet$}}
\put(725,284){\makebox(0,0){$\bullet$}}
\put(719,276){\makebox(0,0){$\bullet$}}
\put(712,268){\makebox(0,0){$\bullet$}}
\put(705,260){\makebox(0,0){$\bullet$}}
\put(699,253){\makebox(0,0){$\bullet$}}
\put(692,247){\makebox(0,0){$\bullet$}}
\put(685,240){\makebox(0,0){$\bullet$}}
\put(679,234){\makebox(0,0){$\bullet$}}
\put(672,228){\makebox(0,0){$\bullet$}}
\put(665,223){\makebox(0,0){$\bullet$}}
\put(659,217){\makebox(0,0){$\bullet$}}
\put(652,212){\makebox(0,0){$\bullet$}}
\put(645,208){\makebox(0,0){$\bullet$}}
\put(639,203){\makebox(0,0){$\bullet$}}
\put(632,198){\makebox(0,0){$\bullet$}}
\put(625,194){\makebox(0,0){$\bullet$}}
\put(619,190){\makebox(0,0){$\bullet$}}
\put(612,186){\makebox(0,0){$\bullet$}}
\put(605,183){\makebox(0,0){$\bullet$}}
\put(599,179){\makebox(0,0){$\bullet$}}
\put(592,176){\makebox(0,0){$\bullet$}}
\put(585,173){\makebox(0,0){$\bullet$}}
\put(578,170){\makebox(0,0){$\bullet$}}
\put(572,167){\makebox(0,0){$\bullet$}}
\put(565,164){\makebox(0,0){$\bullet$}}
\put(558,162){\makebox(0,0){$\bullet$}}
\put(552,159){\makebox(0,0){$\bullet$}}
\put(545,157){\makebox(0,0){$\bullet$}}
\put(538,155){\makebox(0,0){$\bullet$}}
\put(532,153){\makebox(0,0){$\bullet$}}
\put(525,151){\makebox(0,0){$\bullet$}}
\put(518,149){\makebox(0,0){$\bullet$}}
\put(512,147){\makebox(0,0){$\bullet$}}
\put(505,146){\makebox(0,0){$\bullet$}}
\put(498,144){\makebox(0,0){$\bullet$}}
\put(492,143){\makebox(0,0){$\bullet$}}
\put(485,142){\makebox(0,0){$\bullet$}}
\put(478,141){\makebox(0,0){$\bullet$}}
\put(472,139){\makebox(0,0){$\bullet$}}
\put(772,306){\usebox{\plotpoint}}
\put(772.00,306.00){\usebox{\plotpoint}}
\put(751.68,301.95){\usebox{\plotpoint}}
\put(731.41,297.83){\usebox{\plotpoint}}
\put(711.30,292.80){\usebox{\plotpoint}}
\put(691.20,287.77){\usebox{\plotpoint}}
\put(671.35,281.72){\usebox{\plotpoint}}
\put(653.19,271.85){\usebox{\plotpoint}}
\put(639.80,256.07){\usebox{\plotpoint}}
\put(624.79,241.79){\usebox{\plotpoint}}
\put(609.28,228.05){\usebox{\plotpoint}}
\put(592.74,215.53){\usebox{\plotpoint}}
\put(575.35,204.24){\usebox{\plotpoint}}
\put(557.50,193.67){\usebox{\plotpoint}}
\put(539.00,184.43){\usebox{\plotpoint}}
\put(520.55,175.09){\usebox{\plotpoint}}
\put(501.33,167.43){\usebox{\plotpoint}}
\put(481.98,160.14){\usebox{\plotpoint}}
\put(462.47,153.28){\usebox{\plotpoint}}
\put(458,152){\usebox{\plotpoint}}
\put(772,306){\makebox(0,0){$\circ$}}
\put(766,305){\makebox(0,0){$\circ$}}
\put(759,303){\makebox(0,0){$\circ$}}
\put(752,302){\makebox(0,0){$\circ$}}
\put(745,301){\makebox(0,0){$\circ$}}
\put(739,299){\makebox(0,0){$\circ$}}
\put(732,298){\makebox(0,0){$\circ$}}
\put(725,296){\makebox(0,0){$\circ$}}
\put(719,295){\makebox(0,0){$\circ$}}
\put(712,293){\makebox(0,0){$\circ$}}
\put(705,291){\makebox(0,0){$\circ$}}
\put(699,290){\makebox(0,0){$\circ$}}
\put(692,288){\makebox(0,0){$\circ$}}
\put(685,286){\makebox(0,0){$\circ$}}
\put(679,284){\makebox(0,0){$\circ$}}
\put(672,282){\makebox(0,0){$\circ$}}
\put(665,279){\makebox(0,0){$\circ$}}
\put(659,276){\makebox(0,0){$\circ$}}
\put(652,271){\makebox(0,0){$\circ$}}
\put(645,263){\makebox(0,0){$\circ$}}
\put(639,255){\makebox(0,0){$\circ$}}
\put(632,248){\makebox(0,0){$\circ$}}
\put(625,242){\makebox(0,0){$\circ$}}
\put(619,236){\makebox(0,0){$\circ$}}
\put(612,230){\makebox(0,0){$\circ$}}
\put(605,225){\makebox(0,0){$\circ$}}
\put(599,220){\makebox(0,0){$\circ$}}
\put(592,215){\makebox(0,0){$\circ$}}
\put(585,210){\makebox(0,0){$\circ$}}
\put(578,206){\makebox(0,0){$\circ$}}
\put(572,202){\makebox(0,0){$\circ$}}
\put(565,198){\makebox(0,0){$\circ$}}
\put(558,194){\makebox(0,0){$\circ$}}
\put(552,190){\makebox(0,0){$\circ$}}
\put(545,187){\makebox(0,0){$\circ$}}
\put(538,184){\makebox(0,0){$\circ$}}
\put(532,180){\makebox(0,0){$\circ$}}
\put(525,177){\makebox(0,0){$\circ$}}
\put(518,174){\makebox(0,0){$\circ$}}
\put(512,171){\makebox(0,0){$\circ$}}
\put(505,169){\makebox(0,0){$\circ$}}
\put(498,166){\makebox(0,0){$\circ$}}
\put(492,163){\makebox(0,0){$\circ$}}
\put(485,161){\makebox(0,0){$\circ$}}
\put(478,159){\makebox(0,0){$\circ$}}
\put(472,156){\makebox(0,0){$\circ$}}
\put(465,154){\makebox(0,0){$\circ$}}
\put(458,152){\makebox(0,0){$\circ$}}
\put(171.0,131.0){\rule[-0.200pt]{0.400pt}{67.211pt}}
\put(171.0,131.0){\rule[-0.200pt]{160.921pt}{0.400pt}}
\put(839.0,131.0){\rule[-0.200pt]{0.400pt}{67.211pt}}
\put(171.0,410.0){\rule[-0.200pt]{160.921pt}{0.400pt}}
\end{picture}

\begin{picture}(900,450)(0,0)
\put(171.0,131.0){\rule[-0.200pt]{4.818pt}{0.400pt}}
\put(151,131){\makebox(0,0)[r]{ 0}}
\put(819.0,131.0){\rule[-0.200pt]{4.818pt}{0.400pt}}
\put(171.0,171.0){\rule[-0.200pt]{4.818pt}{0.400pt}}
\put(151,171){\makebox(0,0)[r]{ 0.1}}
\put(819.0,171.0){\rule[-0.200pt]{4.818pt}{0.400pt}}
\put(171.0,211.0){\rule[-0.200pt]{4.818pt}{0.400pt}}
\put(151,211){\makebox(0,0)[r]{ 0.2}}
\put(819.0,211.0){\rule[-0.200pt]{4.818pt}{0.400pt}}
\put(171.0,251.0){\rule[-0.200pt]{4.818pt}{0.400pt}}
\put(151,251){\makebox(0,0)[r]{ 0.3}}
\put(819.0,251.0){\rule[-0.200pt]{4.818pt}{0.400pt}}
\put(171.0,290.0){\rule[-0.200pt]{4.818pt}{0.400pt}}
\put(151,290){\makebox(0,0)[r]{ 0.4}}
\put(819.0,290.0){\rule[-0.200pt]{4.818pt}{0.400pt}}
\put(171.0,330.0){\rule[-0.200pt]{4.818pt}{0.400pt}}
\put(151,330){\makebox(0,0)[r]{ 0.5}}
\put(819.0,330.0){\rule[-0.200pt]{4.818pt}{0.400pt}}
\put(171.0,370.0){\rule[-0.200pt]{4.818pt}{0.400pt}}
\put(151,370){\makebox(0,0)[r]{ 0.6}}
\put(819.0,370.0){\rule[-0.200pt]{4.818pt}{0.400pt}}
\put(171.0,410.0){\rule[-0.200pt]{4.818pt}{0.400pt}}
\put(151,410){\makebox(0,0)[r]{ 0.7}}
\put(819.0,410.0){\rule[-0.200pt]{4.818pt}{0.400pt}}
\put(171.0,131.0){\rule[-0.200pt]{0.400pt}{4.818pt}}
\put(171,90){\makebox(0,0){ 0}}
\put(171.0,390.0){\rule[-0.200pt]{0.400pt}{4.818pt}}
\put(305.0,131.0){\rule[-0.200pt]{0.400pt}{4.818pt}}
\put(305,90){\makebox(0,0){ 1}}
\put(305.0,390.0){\rule[-0.200pt]{0.400pt}{4.818pt}}
\put(438.0,131.0){\rule[-0.200pt]{0.400pt}{4.818pt}}
\put(438,90){\makebox(0,0){ 2}}
\put(438.0,390.0){\rule[-0.200pt]{0.400pt}{4.818pt}}
\put(572.0,131.0){\rule[-0.200pt]{0.400pt}{4.818pt}}
\put(572,90){\makebox(0,0){ 3}}
\put(572.0,390.0){\rule[-0.200pt]{0.400pt}{4.818pt}}
\put(705.0,131.0){\rule[-0.200pt]{0.400pt}{4.818pt}}
\put(705,90){\makebox(0,0){ 4}}
\put(705.0,390.0){\rule[-0.200pt]{0.400pt}{4.818pt}}
\put(839.0,131.0){\rule[-0.200pt]{0.400pt}{4.818pt}}
\put(839,90){\makebox(0,0){ 5}}
\put(839.0,390.0){\rule[-0.200pt]{0.400pt}{4.818pt}}
\put(171.0,131.0){\rule[-0.200pt]{0.400pt}{67.211pt}}
\put(171.0,131.0){\rule[-0.200pt]{160.921pt}{0.400pt}}
\put(839.0,131.0){\rule[-0.200pt]{0.400pt}{67.211pt}}
\put(171.0,410.0){\rule[-0.200pt]{160.921pt}{0.400pt}}
\put(30,270){\makebox(0,0){$C$}}
\put(220,350){\makebox(0,0){(e)}}
\put(505,29){\makebox(0,0){$T$}}
\put(632,249){\usebox{\plotpoint}}
\put(619,249.17){\rule{1.300pt}{0.400pt}}
\multiput(622.30,248.17)(-3.302,2.000){2}{\rule{0.650pt}{0.400pt}}
\put(612,250.67){\rule{1.686pt}{0.400pt}}
\multiput(615.50,250.17)(-3.500,1.000){2}{\rule{0.843pt}{0.400pt}}
\put(605,251.67){\rule{1.686pt}{0.400pt}}
\multiput(608.50,251.17)(-3.500,1.000){2}{\rule{0.843pt}{0.400pt}}
\put(599,253.17){\rule{1.300pt}{0.400pt}}
\multiput(602.30,252.17)(-3.302,2.000){2}{\rule{0.650pt}{0.400pt}}
\put(592,255.17){\rule{1.500pt}{0.400pt}}
\multiput(595.89,254.17)(-3.887,2.000){2}{\rule{0.750pt}{0.400pt}}
\put(585,256.67){\rule{1.686pt}{0.400pt}}
\multiput(588.50,256.17)(-3.500,1.000){2}{\rule{0.843pt}{0.400pt}}
\put(578,257.67){\rule{1.686pt}{0.400pt}}
\multiput(581.50,257.17)(-3.500,1.000){2}{\rule{0.843pt}{0.400pt}}
\put(572,259.17){\rule{1.300pt}{0.400pt}}
\multiput(575.30,258.17)(-3.302,2.000){2}{\rule{0.650pt}{0.400pt}}
\put(565,260.67){\rule{1.686pt}{0.400pt}}
\multiput(568.50,260.17)(-3.500,1.000){2}{\rule{0.843pt}{0.400pt}}
\multiput(560.71,262.61)(-1.355,0.447){3}{\rule{1.033pt}{0.108pt}}
\multiput(562.86,261.17)(-4.855,3.000){2}{\rule{0.517pt}{0.400pt}}
\multiput(555.09,265.60)(-0.774,0.468){5}{\rule{0.700pt}{0.113pt}}
\multiput(556.55,264.17)(-4.547,4.000){2}{\rule{0.350pt}{0.400pt}}
\multiput(548.68,269.60)(-0.920,0.468){5}{\rule{0.800pt}{0.113pt}}
\multiput(550.34,268.17)(-5.340,4.000){2}{\rule{0.400pt}{0.400pt}}
\multiput(542.65,273.59)(-0.581,0.482){9}{\rule{0.567pt}{0.116pt}}
\multiput(543.82,272.17)(-5.824,6.000){2}{\rule{0.283pt}{0.400pt}}
\multiput(536.93,279.00)(-0.482,1.304){9}{\rule{0.116pt}{1.100pt}}
\multiput(537.17,279.00)(-6.000,12.717){2}{\rule{0.400pt}{0.550pt}}
\multiput(530.93,294.00)(-0.485,1.560){11}{\rule{0.117pt}{1.300pt}}
\multiput(531.17,294.00)(-7.000,18.302){2}{\rule{0.400pt}{0.650pt}}
\multiput(523.93,315.00)(-0.485,2.552){11}{\rule{0.117pt}{2.043pt}}
\multiput(524.17,315.00)(-7.000,29.760){2}{\rule{0.400pt}{1.021pt}}
\multiput(516.93,349.00)(-0.482,2.118){9}{\rule{0.116pt}{1.700pt}}
\multiput(517.17,349.00)(-6.000,20.472){2}{\rule{0.400pt}{0.850pt}}
\put(505,371.17){\rule{1.500pt}{0.400pt}}
\multiput(508.89,372.17)(-3.887,-2.000){2}{\rule{0.750pt}{0.400pt}}
\multiput(503.93,366.32)(-0.485,-1.332){11}{\rule{0.117pt}{1.129pt}}
\multiput(504.17,368.66)(-7.000,-15.658){2}{\rule{0.400pt}{0.564pt}}
\multiput(496.93,345.67)(-0.482,-2.208){9}{\rule{0.116pt}{1.767pt}}
\multiput(497.17,349.33)(-6.000,-21.333){2}{\rule{0.400pt}{0.883pt}}
\multiput(490.93,324.26)(-0.485,-1.026){11}{\rule{0.117pt}{0.900pt}}
\multiput(491.17,326.13)(-7.000,-12.132){2}{\rule{0.400pt}{0.450pt}}
\multiput(483.93,311.21)(-0.485,-0.721){11}{\rule{0.117pt}{0.671pt}}
\multiput(484.17,312.61)(-7.000,-8.606){2}{\rule{0.400pt}{0.336pt}}
\multiput(476.93,300.54)(-0.482,-0.943){9}{\rule{0.116pt}{0.833pt}}
\multiput(477.17,302.27)(-6.000,-9.270){2}{\rule{0.400pt}{0.417pt}}
\multiput(470.93,290.69)(-0.485,-0.569){11}{\rule{0.117pt}{0.557pt}}
\multiput(471.17,291.84)(-7.000,-6.844){2}{\rule{0.400pt}{0.279pt}}
\multiput(463.93,282.45)(-0.485,-0.645){11}{\rule{0.117pt}{0.614pt}}
\multiput(464.17,283.73)(-7.000,-7.725){2}{\rule{0.400pt}{0.307pt}}
\multiput(456.93,273.37)(-0.482,-0.671){9}{\rule{0.116pt}{0.633pt}}
\multiput(457.17,274.69)(-6.000,-6.685){2}{\rule{0.400pt}{0.317pt}}
\multiput(449.92,266.93)(-0.492,-0.485){11}{\rule{0.500pt}{0.117pt}}
\multiput(450.96,267.17)(-5.962,-7.000){2}{\rule{0.250pt}{0.400pt}}
\multiput(442.65,259.93)(-0.581,-0.482){9}{\rule{0.567pt}{0.116pt}}
\multiput(443.82,260.17)(-5.824,-6.000){2}{\rule{0.283pt}{0.400pt}}
\multiput(435.59,253.93)(-0.599,-0.477){7}{\rule{0.580pt}{0.115pt}}
\multiput(436.80,254.17)(-4.796,-5.000){2}{\rule{0.290pt}{0.400pt}}
\multiput(429.26,248.93)(-0.710,-0.477){7}{\rule{0.660pt}{0.115pt}}
\multiput(430.63,249.17)(-5.630,-5.000){2}{\rule{0.330pt}{0.400pt}}
\multiput(421.68,243.94)(-0.920,-0.468){5}{\rule{0.800pt}{0.113pt}}
\multiput(423.34,244.17)(-5.340,-4.000){2}{\rule{0.400pt}{0.400pt}}
\multiput(414.68,239.94)(-0.920,-0.468){5}{\rule{0.800pt}{0.113pt}}
\multiput(416.34,240.17)(-5.340,-4.000){2}{\rule{0.400pt}{0.400pt}}
\put(405,235.17){\rule{1.300pt}{0.400pt}}
\multiput(408.30,236.17)(-3.302,-2.000){2}{\rule{0.650pt}{0.400pt}}
\multiput(401.68,233.94)(-0.920,-0.468){5}{\rule{0.800pt}{0.113pt}}
\multiput(403.34,234.17)(-5.340,-4.000){2}{\rule{0.400pt}{0.400pt}}
\put(391,229.17){\rule{1.500pt}{0.400pt}}
\multiput(394.89,230.17)(-3.887,-2.000){2}{\rule{0.750pt}{0.400pt}}
\put(385,227.67){\rule{1.445pt}{0.400pt}}
\multiput(388.00,228.17)(-3.000,-1.000){2}{\rule{0.723pt}{0.400pt}}
\put(625.0,249.0){\rule[-0.200pt]{1.686pt}{0.400pt}}
\put(371,227.67){\rule{1.686pt}{0.400pt}}
\multiput(374.50,227.17)(-3.500,1.000){2}{\rule{0.843pt}{0.400pt}}
\put(365,228.67){\rule{1.445pt}{0.400pt}}
\multiput(368.00,228.17)(-3.000,1.000){2}{\rule{0.723pt}{0.400pt}}
\put(358,229.67){\rule{1.686pt}{0.400pt}}
\multiput(361.50,229.17)(-3.500,1.000){2}{\rule{0.843pt}{0.400pt}}
\multiput(353.71,231.61)(-1.355,0.447){3}{\rule{1.033pt}{0.108pt}}
\multiput(355.86,230.17)(-4.855,3.000){2}{\rule{0.517pt}{0.400pt}}
\put(345,234.17){\rule{1.300pt}{0.400pt}}
\multiput(348.30,233.17)(-3.302,2.000){2}{\rule{0.650pt}{0.400pt}}
\put(338,236.17){\rule{1.500pt}{0.400pt}}
\multiput(341.89,235.17)(-3.887,2.000){2}{\rule{0.750pt}{0.400pt}}
\put(331,237.67){\rule{1.686pt}{0.400pt}}
\multiput(334.50,237.17)(-3.500,1.000){2}{\rule{0.843pt}{0.400pt}}
\put(325,237.67){\rule{1.445pt}{0.400pt}}
\multiput(328.00,238.17)(-3.000,-1.000){2}{\rule{0.723pt}{0.400pt}}
\multiput(322.26,236.93)(-0.710,-0.477){7}{\rule{0.660pt}{0.115pt}}
\multiput(323.63,237.17)(-5.630,-5.000){2}{\rule{0.330pt}{0.400pt}}
\multiput(315.92,231.93)(-0.492,-0.485){11}{\rule{0.500pt}{0.117pt}}
\multiput(316.96,232.17)(-5.962,-7.000){2}{\rule{0.250pt}{0.400pt}}
\multiput(309.93,222.54)(-0.482,-0.943){9}{\rule{0.116pt}{0.833pt}}
\multiput(310.17,224.27)(-6.000,-9.270){2}{\rule{0.400pt}{0.417pt}}
\multiput(303.93,211.26)(-0.485,-1.026){11}{\rule{0.117pt}{0.900pt}}
\multiput(304.17,213.13)(-7.000,-12.132){2}{\rule{0.400pt}{0.450pt}}
\multiput(296.93,197.50)(-0.485,-0.950){11}{\rule{0.117pt}{0.843pt}}
\multiput(297.17,199.25)(-7.000,-11.251){2}{\rule{0.400pt}{0.421pt}}
\multiput(289.93,183.99)(-0.482,-1.123){9}{\rule{0.116pt}{0.967pt}}
\multiput(290.17,185.99)(-6.000,-10.994){2}{\rule{0.400pt}{0.483pt}}
\multiput(283.93,171.74)(-0.485,-0.874){11}{\rule{0.117pt}{0.786pt}}
\multiput(284.17,173.37)(-7.000,-10.369){2}{\rule{0.400pt}{0.393pt}}
\multiput(276.93,160.45)(-0.485,-0.645){11}{\rule{0.117pt}{0.614pt}}
\multiput(277.17,161.73)(-7.000,-7.725){2}{\rule{0.400pt}{0.307pt}}
\multiput(269.93,151.37)(-0.482,-0.671){9}{\rule{0.116pt}{0.633pt}}
\multiput(270.17,152.69)(-6.000,-6.685){2}{\rule{0.400pt}{0.317pt}}
\put(632,249){\makebox(0,0){$\bullet$}}
\put(625,249){\makebox(0,0){$\bullet$}}
\put(619,251){\makebox(0,0){$\bullet$}}
\put(612,252){\makebox(0,0){$\bullet$}}
\put(605,253){\makebox(0,0){$\bullet$}}
\put(599,255){\makebox(0,0){$\bullet$}}
\put(592,257){\makebox(0,0){$\bullet$}}
\put(585,258){\makebox(0,0){$\bullet$}}
\put(578,259){\makebox(0,0){$\bullet$}}
\put(572,261){\makebox(0,0){$\bullet$}}
\put(565,262){\makebox(0,0){$\bullet$}}
\put(558,265){\makebox(0,0){$\bullet$}}
\put(552,269){\makebox(0,0){$\bullet$}}
\put(545,273){\makebox(0,0){$\bullet$}}
\put(538,279){\makebox(0,0){$\bullet$}}
\put(532,294){\makebox(0,0){$\bullet$}}
\put(525,315){\makebox(0,0){$\bullet$}}
\put(518,349){\makebox(0,0){$\bullet$}}
\put(512,373){\makebox(0,0){$\bullet$}}
\put(505,371){\makebox(0,0){$\bullet$}}
\put(498,353){\makebox(0,0){$\bullet$}}
\put(492,328){\makebox(0,0){$\bullet$}}
\put(485,314){\makebox(0,0){$\bullet$}}
\put(478,304){\makebox(0,0){$\bullet$}}
\put(472,293){\makebox(0,0){$\bullet$}}
\put(465,285){\makebox(0,0){$\bullet$}}
\put(458,276){\makebox(0,0){$\bullet$}}
\put(452,268){\makebox(0,0){$\bullet$}}
\put(445,261){\makebox(0,0){$\bullet$}}
\put(438,255){\makebox(0,0){$\bullet$}}
\put(432,250){\makebox(0,0){$\bullet$}}
\put(425,245){\makebox(0,0){$\bullet$}}
\put(418,241){\makebox(0,0){$\bullet$}}
\put(411,237){\makebox(0,0){$\bullet$}}
\put(405,235){\makebox(0,0){$\bullet$}}
\put(398,231){\makebox(0,0){$\bullet$}}
\put(391,229){\makebox(0,0){$\bullet$}}
\put(385,228){\makebox(0,0){$\bullet$}}
\put(378,228){\makebox(0,0){$\bullet$}}
\put(371,229){\makebox(0,0){$\bullet$}}
\put(365,230){\makebox(0,0){$\bullet$}}
\put(358,231){\makebox(0,0){$\bullet$}}
\put(351,234){\makebox(0,0){$\bullet$}}
\put(345,236){\makebox(0,0){$\bullet$}}
\put(338,238){\makebox(0,0){$\bullet$}}
\put(331,239){\makebox(0,0){$\bullet$}}
\put(325,238){\makebox(0,0){$\bullet$}}
\put(318,233){\makebox(0,0){$\bullet$}}
\put(311,226){\makebox(0,0){$\bullet$}}
\put(305,215){\makebox(0,0){$\bullet$}}
\put(298,201){\makebox(0,0){$\bullet$}}
\put(291,188){\makebox(0,0){$\bullet$}}
\put(285,175){\makebox(0,0){$\bullet$}}
\put(278,163){\makebox(0,0){$\bullet$}}
\put(271,154){\makebox(0,0){$\bullet$}}
\put(265,146){\makebox(0,0){$\bullet$}}
\put(378.0,228.0){\rule[-0.200pt]{1.686pt}{0.400pt}}
\put(565,261){\usebox{\plotpoint}}
\put(565.00,261.00){\usebox{\plotpoint}}
\put(544.87,264.04){\usebox{\plotpoint}}
\put(525.08,269.97){\usebox{\plotpoint}}
\put(508.65,281.79){\usebox{\plotpoint}}
\put(496.36,297.91){\usebox{\plotpoint}}
\put(488.89,317.22){\usebox{\plotpoint}}
\put(474.75,322.21){\usebox{\plotpoint}}
\put(466.54,303.40){\usebox{\plotpoint}}
\put(456.00,285.67){\usebox{\plotpoint}}
\put(444.53,268.39){\usebox{\plotpoint}}
\put(431.26,252.47){\usebox{\plotpoint}}
\put(415.64,238.97){\usebox{\plotpoint}}
\put(400.47,225.71){\usebox{\plotpoint}}
\put(383.46,214.12){\usebox{\plotpoint}}
\put(364.87,204.95){\usebox{\plotpoint}}
\put(345.45,198.22){\usebox{\plotpoint}}
\put(325.72,201.76){\usebox{\plotpoint}}
\put(309.90,215.10){\usebox{\plotpoint}}
\put(292.45,225.58){\usebox{\plotpoint}}
\put(281.23,209.46){\usebox{\plotpoint}}
\put(272.85,190.49){\usebox{\plotpoint}}
\put(265.97,170.92){\usebox{\plotpoint}}
\put(265,168){\usebox{\plotpoint}}
\put(565,261){\makebox(0,0){$\circ$}}
\put(558,261){\makebox(0,0){$\circ$}}
\put(552,261){\makebox(0,0){$\circ$}}
\put(545,264){\makebox(0,0){$\circ$}}
\put(538,266){\makebox(0,0){$\circ$}}
\put(532,267){\makebox(0,0){$\circ$}}
\put(525,270){\makebox(0,0){$\circ$}}
\put(518,273){\makebox(0,0){$\circ$}}
\put(512,277){\makebox(0,0){$\circ$}}
\put(505,287){\makebox(0,0){$\circ$}}
\put(498,293){\makebox(0,0){$\circ$}}
\put(492,311){\makebox(0,0){$\circ$}}
\put(485,325){\makebox(0,0){$\circ$}}
\put(478,326){\makebox(0,0){$\circ$}}
\put(472,319){\makebox(0,0){$\circ$}}
\put(465,299){\makebox(0,0){$\circ$}}
\put(458,289){\makebox(0,0){$\circ$}}
\put(452,279){\makebox(0,0){$\circ$}}
\put(445,269){\makebox(0,0){$\circ$}}
\put(438,260){\makebox(0,0){$\circ$}}
\put(432,253){\makebox(0,0){$\circ$}}
\put(425,248){\makebox(0,0){$\circ$}}
\put(418,242){\makebox(0,0){$\circ$}}
\put(411,233){\makebox(0,0){$\circ$}}
\put(405,227){\makebox(0,0){$\circ$}}
\put(398,225){\makebox(0,0){$\circ$}}
\put(391,220){\makebox(0,0){$\circ$}}
\put(385,215){\makebox(0,0){$\circ$}}
\put(378,211){\makebox(0,0){$\circ$}}
\put(371,208){\makebox(0,0){$\circ$}}
\put(365,205){\makebox(0,0){$\circ$}}
\put(358,202){\makebox(0,0){$\circ$}}
\put(351,201){\makebox(0,0){$\circ$}}
\put(345,198){\makebox(0,0){$\circ$}}
\put(338,197){\makebox(0,0){$\circ$}}
\put(331,200){\makebox(0,0){$\circ$}}
\put(325,202){\makebox(0,0){$\circ$}}
\put(318,208){\makebox(0,0){$\circ$}}
\put(311,214){\makebox(0,0){$\circ$}}
\put(305,220){\makebox(0,0){$\circ$}}
\put(298,224){\makebox(0,0){$\circ$}}
\put(291,226){\makebox(0,0){$\circ$}}
\put(285,217){\makebox(0,0){$\circ$}}
\put(278,203){\makebox(0,0){$\circ$}}
\put(271,186){\makebox(0,0){$\circ$}}
\put(265,168){\makebox(0,0){$\circ$}}
\put(171.0,131.0){\rule[-0.200pt]{0.400pt}{67.211pt}}
\put(171.0,131.0){\rule[-0.200pt]{160.921pt}{0.400pt}}
\put(839.0,131.0){\rule[-0.200pt]{0.400pt}{67.211pt}}
\put(171.0,410.0){\rule[-0.200pt]{160.921pt}{0.400pt}}
\end{picture}

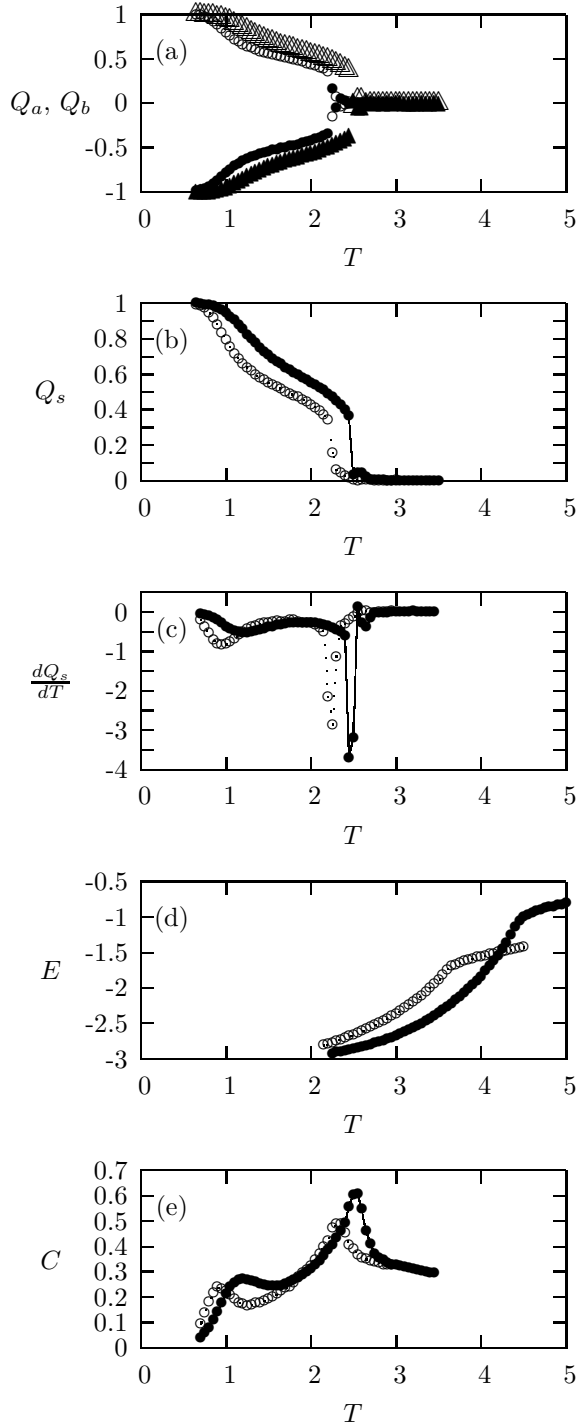
\noindent {\bf Fig-3.} Temperature dependences of all dynamic quantities.
(a) $Q_a (\circ)$ and $Q_b (\bullet)$ for $h_0=4.0$ and $Q_a (\triangle)$ and
$Q_b (\blacktriangle)$ for $h_0=3.9$. (b) $Q_s$ versus $T$ for $h_0 = 4.0
(\circ)$ and $h_0 = 3.9 (\bullet)$. (c) ${{dQ_s} \over {dT}}$ versus $T$, for
$h_0 = 4.0 (\circ)$ and $h_0 = 3.9 (\bullet)$. (d) $E$ versus $T$, for
$h_0 = 4.0 (\circ)$ and $h_0 = 3.9 (\bullet)$. (e) $C = {{dE} \over {dT}}$ 
versus $T$, for $h_0 = 4.0 (\circ)$ and $h_0 = 3.9 (\bullet)$.
Here, $f=0.01$ and $\lambda = 5$.

\newpage

% GNUPLOT: LaTeX picture FIG-4
\setlength{\unitlength}{0.240900pt}
\ifx\plotpoint\undefined\newsavebox{\plotpoint}\fi
\sbox{\plotpoint}{\rule[-0.200pt]{0.400pt}{0.400pt}}%
\begin{picture}(1050,900)(0,0)
\sbox{\plotpoint}{\rule[-0.200pt]{0.400pt}{0.400pt}}%
\put(131.0,131.0){\rule[-0.200pt]{4.818pt}{0.400pt}}
\put(111,131){\makebox(0,0)[r]{ 0}}
\put(969.0,131.0){\rule[-0.200pt]{4.818pt}{0.400pt}}
\put(131.0,277.0){\rule[-0.200pt]{4.818pt}{0.400pt}}
\put(111,277){\makebox(0,0)[r]{ 1}}
\put(969.0,277.0){\rule[-0.200pt]{4.818pt}{0.400pt}}
\put(131.0,422.0){\rule[-0.200pt]{4.818pt}{0.400pt}}
\put(111,422){\makebox(0,0)[r]{ 2}}
\put(969.0,422.0){\rule[-0.200pt]{4.818pt}{0.400pt}}
\put(131.0,568.0){\rule[-0.200pt]{4.818pt}{0.400pt}}
\put(111,568){\makebox(0,0)[r]{ 3}}
\put(969.0,568.0){\rule[-0.200pt]{4.818pt}{0.400pt}}
\put(131.0,713.0){\rule[-0.200pt]{4.818pt}{0.400pt}}
\put(111,713){\makebox(0,0)[r]{ 4}}
\put(969.0,713.0){\rule[-0.200pt]{4.818pt}{0.400pt}}
\put(131.0,859.0){\rule[-0.200pt]{4.818pt}{0.400pt}}
\put(111,859){\makebox(0,0)[r]{ 5}}
\put(969.0,859.0){\rule[-0.200pt]{4.818pt}{0.400pt}}
\put(131.0,131.0){\rule[-0.200pt]{0.400pt}{4.818pt}}
\put(131,90){\makebox(0,0){ 0}}
\put(131.0,839.0){\rule[-0.200pt]{0.400pt}{4.818pt}}
\put(303.0,131.0){\rule[-0.200pt]{0.400pt}{4.818pt}}
\put(303,90){\makebox(0,0){ 1}}
\put(303.0,839.0){\rule[-0.200pt]{0.400pt}{4.818pt}}
\put(474.0,131.0){\rule[-0.200pt]{0.400pt}{4.818pt}}
\put(474,90){\makebox(0,0){ 2}}
\put(474.0,839.0){\rule[-0.200pt]{0.400pt}{4.818pt}}
\put(646.0,131.0){\rule[-0.200pt]{0.400pt}{4.818pt}}
\put(646,90){\makebox(0,0){ 3}}
\put(646.0,839.0){\rule[-0.200pt]{0.400pt}{4.818pt}}
\put(817.0,131.0){\rule[-0.200pt]{0.400pt}{4.818pt}}
\put(817,90){\makebox(0,0){ 4}}
\put(817.0,839.0){\rule[-0.200pt]{0.400pt}{4.818pt}}
\put(989.0,131.0){\rule[-0.200pt]{0.400pt}{4.818pt}}
\put(989,90){\makebox(0,0){ 5}}
\put(989.0,839.0){\rule[-0.200pt]{0.400pt}{4.818pt}}
\put(131.0,131.0){\rule[-0.200pt]{0.400pt}{175.375pt}}
\put(131.0,131.0){\rule[-0.200pt]{206.692pt}{0.400pt}}
\put(989.0,131.0){\rule[-0.200pt]{0.400pt}{175.375pt}}
\put(131.0,859.0){\rule[-0.200pt]{206.692pt}{0.400pt}}
\put(30,495){\makebox(0,0){$h_0$}}
\put(560,29){\makebox(0,0){$T$}}
\put(303,277){\makebox(0,0)[l]{$Q_s \neq 0$}}
\put(732,641){\makebox(0,0)[l]{$Q_s = 0$}}
\sbox{\plotpoint}{\rule[-0.400pt]{0.800pt}{0.800pt}}%
\put(895,167){\makebox(0,0){$\bullet$}}
\put(877,204){\makebox(0,0){$\bullet$}}
\put(877,277){\makebox(0,0){$\bullet$}}
\put(852,349){\makebox(0,0){$\bullet$}}
\put(835,422){\makebox(0,0){$\bullet$}}
\put(792,495){\makebox(0,0){$\bullet$}}
\put(740,568){\makebox(0,0){$\bullet$}}
\put(654,641){\makebox(0,0){$\bullet$}}
\put(637,655){\makebox(0,0){$\bullet$}}
\put(620,662){\makebox(0,0){$\bullet$}}
\put(611,670){\makebox(0,0){$\bullet$}}
\put(594,677){\makebox(0,0){$\bullet$}}
\put(577,684){\makebox(0,0){$\bullet$}}
\put(560,692){\makebox(0,0){$\bullet$}}
\put(551,699){\makebox(0,0){$\bullet$}}
\put(526,706){\makebox(0,0){$\bullet$}}
\put(517,713){\makebox(0,0){$\bullet$}}
\put(466,728){\makebox(0,0){$\bullet$}}
\sbox{\plotpoint}{\rule[-0.200pt]{0.400pt}{0.400pt}}%
\sbox{\plotpoint}{\rule[-0.400pt]{0.800pt}{0.800pt}}%
\put(448,670){\makebox(0,0){$\circ$}}
\put(397,677){\makebox(0,0){$\circ$}}
\put(380,684){\makebox(0,0){$\circ$}}
\put(354,692){\makebox(0,0){$\circ$}}
\put(337,699){\makebox(0,0){$\circ$}}
\put(320,706){\makebox(0,0){$\circ$}}
\put(294,713){\makebox(0,0){$\circ$}}
\put(277,728){\makebox(0,0){$\circ$}}
\sbox{\plotpoint}{\rule[-0.500pt]{1.000pt}{1.000pt}}%
\put(895,167){\makebox(0,0){$\blacktriangle$}}
\put(895,204){\makebox(0,0){$\blacktriangle$}}
\put(877,277){\makebox(0,0){$\blacktriangle$}}
\put(860,349){\makebox(0,0){$\blacktriangle$}}
\put(835,422){\makebox(0,0){$\blacktriangle$}}
\put(792,495){\makebox(0,0){$\blacktriangle$}}
\put(740,568){\makebox(0,0){$\blacktriangle$}}
\put(663,641){\makebox(0,0){$\blacktriangle$}}
\put(646,655){\makebox(0,0){$\blacktriangle$}}
\put(629,662){\makebox(0,0){$\blacktriangle$}}
\put(620,670){\makebox(0,0){$\blacktriangle$}}
\put(603,677){\makebox(0,0){$\blacktriangle$}}
\put(594,684){\makebox(0,0){$\blacktriangle$}}
\put(586,692){\makebox(0,0){$\blacktriangle$}}
\put(569,699){\makebox(0,0){$\blacktriangle$}}
\put(551,706){\makebox(0,0){$\blacktriangle$}}
\put(526,713){\makebox(0,0){$\blacktriangle$}}
\put(483,728){\makebox(0,0){$\blacktriangle$}}
\put(440,670){\makebox(0,0){$\triangle$}}
\put(397,677){\makebox(0,0){$\triangle$}}
\put(388,684){\makebox(0,0){$\triangle$}}
\put(354,692){\makebox(0,0){$\triangle$}}
\put(337,699){\makebox(0,0){$\triangle$}}
\put(320,706){\makebox(0,0){$\triangle$}}
\put(285,713){\makebox(0,0){$\triangle$}}
\put(243,728){\makebox(0,0){$\triangle$}}
\sbox{\plotpoint}{\rule[-0.200pt]{0.400pt}{0.400pt}}%
\put(131.0,131.0){\rule[-0.200pt]{0.400pt}{175.375pt}}
\put(131.0,131.0){\rule[-0.200pt]{206.692pt}{0.400pt}}
\put(989.0,131.0){\rule[-0.200pt]{0.400pt}{175.375pt}}
\put(131.0,859.0){\rule[-0.200pt]{206.692pt}{0.400pt}}
\end{picture}

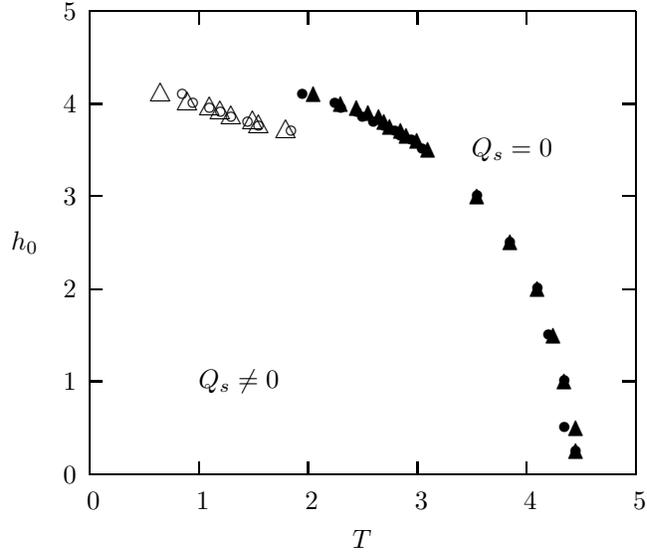
\noindent {\bf Fig-4.} The phase diagram. Transition temperatures obtained 
from the high temperature peak ($\blacktriangle$) and the low temperature peak
($\triangle$) of dynamic specific heat ($C$). Transition temperatures obtained
from the high temperature dip ($\bullet$) and the low temperature dip ($\circ$)
of ${{dQ_s} \over {dT}}$.
Here, $f=0.01$ and $\lambda = 5$.
\end{document}